\documentclass[10pt]{article}
\usepackage{graphicx}
\usepackage{amsmath}
\usepackage{amssymb}
\usepackage{mathrsfs}
\usepackage{multirow}
\usepackage{caption2}
\usepackage{authblk}
\usepackage[figuresright]{rotating}
\begin{document}
\bibliographystyle{prsty}
\begin{center}
{\large {\bf \sc{Systematic study of the strong decays of the $P_c$ states and their possible isospin cousins via the QCD sum rules}}} \\[2mm]
Xiu-Wu Wang\footnote{E-mail: wangxiuwu2020@163.com.}, Xin Li and
Zhi-Gang  Wang\footnote{E-mail: zgwang@aliyun.com.}\\
 Department of Physics, North China Electric Power University, Baoding 071003, P. R. China\\
\end{center}
\begin{abstract}
In the present work, the strong decays of the discovered $P_c(4380)$, $P_c(4440)$, $P_c(4457)$ and their possible isospin cousins are systematically studied via the assignment that they are the meson-baryon molecular states. In detail, the strong decay constants and partial decay widths of their decay channels are calculated under the framework of QCD sum rules. The decay widths of the discovered $P_c(4380)$, $P_c(4440)$ and $P_c(4457)$ are in good agreement with the experiments. The predictions of the decays of these three related possible isospin cousins are presented which would shed light on their findings in experiments. In return, this may testify to the assignments of the discovered $P_c$ states.
\end{abstract}



\section{Introduction}
So far, the LHCb collaboration has discoveried the following exotic $P_c$ states composed of five valence quarks: the $P_c(4380)$ observed in 2015 via analyzing the $\Lambda_b^0\rightarrow J/\psi K^-p$ decays \cite{RAaij1}, the $P_c(4312)$, $P_c(4440)$ and $P_c(4457)$ found in 2019 in the $J/\psi p$ mass spectrum \cite{RAaij2}, the $P_c(4337)$  in the $J/\psi p$ and $J/\psi \bar{p}$ systems in the $B_s^0\rightarrow J/\psi p\bar{p}$ decays \cite{RAaij3}. It is worth mentioning that the $P_c(4380)$ has neither been confirmed nor conclusively excluded in the subsequent investigations. For the $P_c(4337)$, it is hard to assign it as the molecular state due to lacking nearby meson-baryon thresholds, furthermore, its existence still needs confirmation. As for the strong decay of $P_c(4312)$, it is fully analyzed in Ref. \cite{Myself-DcPc4312} by our group, thus, the strong decays of the observed $P_c(4380)$, $P_c(4440)$ and $P_c(4457)$ are focused in the present work. Need to point out, for the $P_c(4380)$ having a broad width, its existence remains a controversial question, it was neither confirmed nor refuted in the updated analysis. In 2019, the clear signal for this state was not observed in the $J/\psi p$ mass spectrum by the LHCb group \cite{RAaij2}, and therefore its existence needs further experimental confirmation. In the present study, we still treat it as the `$P_c(4380)$' state with the mass and width proposed by LHCb in 2015 \cite{RAaij1} for the systematic study of these discovered $P_c$ states under the meson-baryon molecular picture, moreover, the prediction of the decays for this controversial state derived in the present work may provide a reference for its further experimental confirmation in the future. The Breit-Wigner masses and widths of $P_c(4380)$, $P_c(4440)$ and $P_c(4457)$ are listed in \cite{RAaij1,RAaij2}.
\begin{eqnarray}
\notag && P_c(4380):M=4380\pm8\pm29\,\rm{MeV}\, ,\,\, \,\Gamma=205\pm18\pm86\,\rm{MeV}\,,\\
\notag && P_c(4440):M=4440.3\pm1.3^{+4.1}_{-4.7}\,\rm{MeV}\,, \,\,\,\Gamma=20.6\pm4.9^{+8.7}_{-10.1}\,\rm{MeV}\,,\\
&& P_c(4457):M=4457.3\pm0.6^{+4.1}_{-1.7}\,\rm{MeV}\,, \,\,\,\Gamma=6.4\pm2.0^{+5.7}_{-1.9}\,\rm{MeV}\,.
\end{eqnarray}
Via the experiments \cite{RAaij1,RAaij2}, one could determine that the isospin of $P_c(4380)$, $P_c(4440)$ and $P_c(4457)$ is $\frac{1}{2}$. Since the spins and parities of these $P_c$ states have not been definitely determined experimentally yet, theoretical groups have different arguments about their physical natures which will surely deepen our understanding of the non-perturbative behavior of strong interactions,
this is still a hotly debated and open question. Based on the fact that the masses of these $P_c$ states are just below the thresholds of the  $\Sigma_c^{(*)}\bar{D}^{(*)}$ pairs, a typical interpretation is that they are the S-wave hidden-charm meson-baryon molecules \cite{Mass-mole-JunHe,Mass-mole-HXChen-2,Mass-mole-HXChen,mass-mole-WZG,mass-mole-MengLin,Mass-mole-LMeng,
Mass-mole-MingZhu,Mass-JRZhang-EPJC2019,mass-mole-Azizi,CWXiao-Pc-mole1,HXChen-Pc-mole2,MZLiu-Pc-mole3,MZLiu-Pc-mole4,CWXiao-Pc-mole5,Decay-mole-Sakai}. Another typical assignment is that they are the compact pentaquark states \cite{mass-penta-WZG-EPJC-16,mass-penta-WZG-IJMPA,WZG-Pc-Penta2,WZG-Pc-Penta4,AAli-Compact5-Pc1,RZhu-Compact5-Pc2,JFGiron-Compact5-Pc4,FStancu-Compact5-Pc5,JFGiron-Compact5-Pc6}.
In the molecular picture, it is widely accepted that the $J^P$ of $P_c(4312)$ is $\frac{1}{2}^-$. The divergences mainly come from the other $P_c$ states, for example, $P_c(4440)$ and $P_c(4457)$ are assigned as the $\Sigma_c\bar{D}^*$ molecules with $J^P$ being $\frac{1}{2}^-$ and $\frac{3}{2}^-$ in Refs. \cite{Mass-mole-JunHe,Decay-mole-Zou,Decay-mole-HeChen}, respectively. However, in Refs.  \cite{NYalikun-Pc4,MPavon-Pc5,MLDu-Pc6,FZPeng-Pc7}, their $J^P$ are attributed as $\frac{3}{2}^-$ and $\frac{1}{2}^-$, respectively. As for our argument for the $P_c(4312)$, $P_c(4380)$, $P_c(4440)$ and $P_c(4457)$, considering these $P_c$ states were discovered in $J/\psi p$ invariant mass spectrum, their isospins should
be $I=\frac{1}{2}$ via the conservation of the isospins in the strong interactions \cite{mass-mole-WXW-SCPMA}. Thus we constructed currents with high and low isospins for the first time to interpolate the hidden-charm $P_c$ states under the framework of the QCD sum rules to avoid pollution between high and low isospin states. The study nicely assigns the $P_c(4312)$, $P_c(4380)$, $P_c(4440)$ and $P_c(4457)$ as the $\bar{D}\Sigma_c$, $\bar{D}\Sigma_c^*$, $\bar{D}^*\Sigma_c$ and $\bar{D}^*\Sigma_c^*$ molecular states with their $J^P$  being $\frac{1}{2}^-$, $\frac{3}{2}^-$, $\frac{3}{2}^-$ and $\frac{5}{2}^-$, respectively. Moreover, four high isospin cousins of the $P_c(4312)$, $P_c(4380)$, $P_c(4440)$ and $P_c(4457)$ are predicted as the $\bar{D}\Sigma_c$, $\bar{D}\Sigma_c^*$, $\bar{D}^*\Sigma_c$ and $\bar{D}^*\Sigma_c^*$ resonance states. For more insights on these $P_c$ states, one can check the relevant reviews in Refs. \cite{FKGuo-Review1,HXChen-Review2,LMeng-Review3,ZGWang-Review4}. Motivated by these heated debates regarding the nature of these $P_c$ states, we turn our attention to their strong decays and try to find more information to determine their natures.

The mass and width are two basic parameters  to determine the physical nature of a hadronic state. Many theoretical groups have applied different methods to study their masses \cite{Mass-mole-JunHe,Mass-mole-HXChen-2,Mass-mole-HXChen,mass-mole-WZG,mass-mole-MengLin,Mass-mole-LMeng,
Mass-mole-MingZhu,Mass-JRZhang-EPJC2019,mass-mole-Azizi,mass-penta-WZG-EPJC-16,mass-penta-WZG-IJMPA,mass-mole-WXW-SCPMA,mass-mole-WXW-CPC,
mass-penta-WZG-CPC} and strong decays \cite{Decay-mole-Sakai,Decay-mole-Zou,Decay-mole-HeChen,Decay-mole-GJWang,Decay-mole-Gutsche,Decay-mole-WZG-WX,Decay-mole-HuangMQ} under the physical picture of meson-baryon hadronic molecules.
For the strong decay, the isospin, spin and parity should be conserved, different $IJ^P$ result in different decay modes.
In our previous work, the masses of the observed $P_c$ and $P_{cs}$ are studied by clearly differentiating the isospins of the currents interpolating the exotic states for the first time in a comprehensive way \cite{mass-mole-WXW-SCPMA,mass-mole-WXW-IJMPA}. Results show that the mass of the high isospin state is several dozen MeV above that of the low one which is solid proof of the necessity to differentiate the isospins to avoid pollution between the high and low isospins in studying these $P_c$ and $P_{cs}$ states. Hence, the necessity to study the strong decay of the possible high isospin cousins of these $P_c$ and $P_{cs}$ is also obvious. Their observation in the predicted decay modes would shed light on the nature of the $P_c$ states in return. Furthermore, the $J^P$ and pole residues are also derived \cite{mass-mole-WXW-SCPMA,mass-mole-WXW-IJMPA}, ready for the present work to study the strong decays.

The article is arranged as follows: The QCD sum rules for the strong decays of $P_c(4380)$, $P_c(4410)$, $P_c(4440)$, $P_c(4470)$, $P_c(4457)$ and $P_c(4620)$ are derived in Sect. 2. The numerical results and discussions are presented in Sect. 3, and Sect. 4 is reserved for the conclusions of the present study.
\section{QCD sum rules for  strong decays of the pentaquark molecular states}
Following the results derived in Ref. \cite{mass-mole-WXW-SCPMA}, the quantum numbers $(I,J^P)$ of the $P_c(4380)$, $P_c(4440)$ and $P_c(4457)$ are $(\frac{1}{2},\frac{3}{2}^-)$, $(\frac{1}{2},\frac{3}{2}^-)$ and $(\frac{1}{2},\frac{5}{2}^-)$, respectively. The three possible corresponding isospin cousins of $P_c(4380)$, $P_c(4440)$ and $P_c(4457)$ are $P_c(4410)$, $P_c(4470)$ and $P_c(4620)$ with their $(I,J^P)$ quantum numbers assigned as $(\frac{3}{2},\frac{3}{2}^-)$, $(\frac{3}{2},\frac{3}{2}^-)$ and $(\frac{3}{2},\frac{5}{2}^-)$, respectively. The $(I,J^P)$ for $N$, $\Delta$, $\eta_c$, $J/\psi$ are $(\frac{1}{2},\frac{1}{2}^+)$, $(\frac{3}{2},\frac{3}{2}^+)$, $(0,0^-)$ and $(0,1^-)$. Considering the conservation of isospin $I$ in the strong decays, the following decay channels are studied in the present work.
\begin{eqnarray}
P_c(4380)&\rightarrow&\eta_c+N\, , \nonumber\\
P_c(4380)&\rightarrow&J/\psi+N\, , \nonumber\\
P_c(4410)&\rightarrow&\eta_c+\Delta\, , \nonumber\\
P_c(4410)&\rightarrow&J/\psi+\Delta\, , \nonumber\\
P_c(4440)&\rightarrow&\eta_c+N\, , \nonumber\\
P_c(4440)&\rightarrow&J/\psi+N\, ,\nonumber\\
P_c(4470)&\rightarrow&\eta_c+\Delta\, , \nonumber\\
P_c(4470)&\rightarrow&J/\psi+\Delta\, ,\nonumber\\
P_c(4457)&\rightarrow&\eta_c+N\, , \nonumber\\
P_c(4457)&\rightarrow&J/\psi+N\, ,\nonumber\\
P_c(4620)&\rightarrow&\eta_c+\Delta\, , \nonumber\\
P_c(4620)&\rightarrow&J/\psi+\Delta\, ,
\end{eqnarray}
where, the proton is marked as $N$ to avoid confusion with the four-momentum $p_\mu$, the currents $\mathcal{J}_{\eta_c}(x)$, $\mathcal{J}_{J/\psi,\mu}(x)$, $\mathcal{J}_{N}(x)$, $\mathcal{J}_{\Delta,\mu}(x)$, $\mathcal{J}_{P_A,\mu}(x)$, $\mathcal{J}_{P_B,\mu}(x)$, $\mathcal{J}_{P_C,\mu}(x)$, $\mathcal{J}_{P_D,\mu}(x)$, $\mathcal{J}_{P_E,\mu\nu}(x)$ and $\mathcal{J}_{P_F,\mu\nu}(x)$ are applied to  interpolate the $\eta_c$, $J/\psi$, $N$, $\Delta$, $P_c(4380)$, $P_c(4410)$, $P_c(4440)$, $P_c(4470)$, $P_c(4457)$ and $P_c(4620)$, respectively. For convenience, $P_{A\sim F}$ are used to represent the states $P_c(4380)$, $P_c(4410)$, $P_c(4440)$, $P_c(4470)$, $P_c(4457)$ and $P_c(4620)$, respectively. The mentioned currents are expressed as,
\begin{eqnarray}\label{Currents-meson}
\mathcal{J}_{\eta_c}(x) &=& \overline{c}(x)\textsf{i}\gamma_5c(x)\, ,
\end{eqnarray}
\begin{eqnarray}
\mathcal{J}_{J/\psi,\mu}(x) &=& \overline{c}(x)\gamma_\mu c(x)\, ,
\end{eqnarray}
\begin{eqnarray}
\mathcal{J}_{N}(x) &=& \varepsilon^{ijk}u^{iT}(x)\textsc{C}\gamma_{\alpha}u^{j}(x)\gamma^{\alpha}\gamma_5d^{k}(x)\, ,
\end{eqnarray}
\begin{eqnarray}
\mathcal{J}_{\Delta,\mu}(x) &=& \frac{1}{\sqrt{3}} \varepsilon^{ijk}u^{iT}(x)\textsc{C}\gamma_{\mu}u^{j}(x)d^{k}(x)+\sqrt{\frac{2}{3}} \varepsilon^{ijk}u^{iT}(x)\textsc{C}\gamma_{\mu}d^{j}(x)u^{k}(x)\, ,
\end{eqnarray}
\begin{eqnarray}
\mathcal{J}_{P_A,\mu}(x)&=&\frac{1}{\sqrt{3}}\bar{c}(x)\textsf{i}\gamma_5u(x)\varepsilon^{ijk}u^{iT}(x)\textsc{C}\gamma_{\mu}d^j(x)c^k(x)\nonumber\\
&&-\sqrt{\frac{2}{3}}\bar{c}(x)\textsf{i}\gamma_5d(x)\varepsilon^{ijk}u^{iT}(x)\textsc{C}\gamma_{\mu}u^j(x)c^k(x)\, ,
\end{eqnarray}
\begin{eqnarray}
\mathcal{J}_{P_B,\mu}(x)&=&\sqrt{\frac{2}{3}}\bar{c}(x)\textsf{i}\gamma_5u(x)\varepsilon^{ijk}u^{iT}(x)\textsc{C}\gamma_{\mu}d^j(x)c^k(x)\nonumber\\
&&+\frac{1}{\sqrt{3}}\bar{c}(x)\textsf{i}\gamma_5d(x)\varepsilon^{ijk}u^{iT}(x)\textsc{C}\gamma_{\mu}u^j(x)c^k(x)\, ,
\end{eqnarray}
\begin{eqnarray}
\mathcal{J}_{P_C,\mu}(x)&=&\frac{1}{\sqrt{3}}\bar{c}(x)\gamma_{\mu} u(x)\varepsilon^{ijk}u^{iT}(x)\textsc{C}\gamma_{\nu}d^j(x)\gamma^{\nu}\gamma_5c^k(x)\nonumber\\
&&-\sqrt{\frac{2}{3}}\bar{c}(x)\gamma_{\mu} d(x)\varepsilon^{ijk}u^{iT}(x)\textsc{C}\gamma_{\nu}u^j(x)\gamma^{\nu}\gamma_5c^k(x)\, ,
\end{eqnarray}
\begin{eqnarray}
\mathcal{J}_{P_D,\mu}(x)&=&\sqrt{\frac{2}{3}}\bar{c}(x)\gamma_{\mu} u(x)\varepsilon^{ijk}u^{iT}(x)\textsc{C}\gamma_{\nu}d^j(x)\gamma^{\nu}\gamma_5c^k(x)\nonumber\\
&&+\frac{1}{\sqrt{3}}\bar{c}(x)\gamma_{\mu} d(x)\varepsilon^{ijk}u^{iT}(x)\textsc{C}\gamma_{\nu}u^j(x)\gamma^{\nu}\gamma_5c^k(x)\, ,
\end{eqnarray}
\begin{eqnarray}
\mathcal{J}_{P_E,\mu\nu}(x)&=&\frac{1}{\sqrt{3}}\bar{c}(x)\gamma_{\mu} u(x)\varepsilon^{ijk}u^{iT}(x)\textsc{C}\gamma_{\nu}d^j(x)c^k(x)\nonumber\\
&&-\sqrt{\frac{2}{3}}\bar{c}(x)\gamma_{\mu} d(x)\varepsilon^{ijk}u^{iT}(x)\textsc{C}\gamma_{\nu}u^j(x)c^k(x)+(\mu\leftrightarrow\nu)\, ,
\end{eqnarray}
\begin{eqnarray}
\mathcal{J}_{P_F,\mu\nu}(x)&=&\sqrt{\frac{2}{3}}\bar{c}(x)\gamma_{\mu} u(x)\varepsilon^{ijk}u^{iT}(x)\textsc{C}\gamma_{\nu}d^j(x)c^k(x)\nonumber\\
&&+\frac{1}{\sqrt{3}}\bar{c}(x)\gamma_{\mu} d(x)\varepsilon^{ijk}u^{iT}(x)\textsc{C}\gamma_{\nu}u^j(x)c^k(x)+(\mu\leftrightarrow\nu)\, ,
\end{eqnarray}
where the $\textsc{C}$ is the charge conjugation matrix, $\varepsilon^{ijk}$ represents the antisymmetric tensor,
and the $i$, $j$ and $k$ are the color indices. Following previous works, the three-point correlation functions in the QCD sum rules are constructed to study the related strong decays.
\begin{eqnarray}\label{PI1}
\Pi_{\mu,1}(p,q) &=& \textsf{i}^2 \int d^4xd^4y e^{\textsf{i}p\cdot x}e^{\textsf{i}q\cdot y} \langle 0|\textsf{T} \left\{\mathcal{J}_{\eta_c}(x)\mathcal{J}_N(y) \bar{\mathcal{J}}_{P_A,\mu}(0) \right\}|0\rangle\, ,
\end{eqnarray}
\begin{eqnarray}\label{PI2}
\Pi_{\mu\zeta,2}(p,q) &=& \textsf{i}^2 \int d^4xd^4y e^{\textsf{i}p\cdot x}e^{\textsf{i}q\cdot y} \langle 0|\textsf{T} \left\{ \mathcal{J}_{J/\psi,\zeta}(x)\mathcal{J}_N(y) \bar{\mathcal{J}}_{P_A,\mu}(0) \right\}|0\rangle\, ,
\end{eqnarray}
\begin{eqnarray}\label{PI3}
\Pi_{\mu\chi,3}(p,q) &=& \textsf{i}^2 \int d^4xd^4y e^{\textsf{i}p\cdot x}e^{\textsf{i}q\cdot y} \langle 0|\textsf{T} \left\{ \mathcal{J}_{\eta_c}(x)\mathcal{J}_{\Delta,\chi}(y) \bar{\mathcal{J}}_{P_B,\mu}(0) \right\}|0\rangle\, ,
\end{eqnarray}
\begin{eqnarray}\label{PI4}
\Pi_{\mu\chi\zeta,4}(p,q) &=& \textsf{i}^2 \int d^4xd^4y e^{\textsf{i}p\cdot x}e^{\textsf{i}q\cdot y} \langle 0|\textsf{T} \left\{ \mathcal{J}_{J/\psi,\zeta}(x)\mathcal{J}_{\Delta,\chi}(y) \bar{\mathcal{J}}_{P_B,\mu}(0) \right\}|0\rangle\, .
\end{eqnarray}
\begin{eqnarray}\label{PI5}
\Pi_{\mu,5}(p,q) &=& \textsf{i}^2 \int d^4xd^4y e^{\textsf{i}p\cdot x}e^{\textsf{i}q\cdot y} \langle 0|\textsf{T} \left\{ \mathcal{J}_{\eta_c}(x)\mathcal{J}_N(y) \bar{\mathcal{J}}_{P_C,\mu}(0) \right\}|0\rangle\, ,
\end{eqnarray}
\begin{eqnarray}\label{PI6}
\Pi_{\mu\zeta,6}(p,q) &=& \textsf{i}^2 \int d^4xd^4y e^{\textsf{i}p\cdot x}e^{\textsf{i}q\cdot y} \langle 0|\textsf{T} \left\{ \mathcal{J}_{J/\psi,\zeta}(x)\mathcal{J}_N(y) \bar{\mathcal{J}}_{P_C,\mu}(0) \right\}|0\rangle\, ,
\end{eqnarray}
\begin{eqnarray}\label{PI7}
\Pi_{\mu\chi,7}(p,q) &=& \textsf{i}^2 \int d^4xd^4y e^{\textsf{i}p\cdot x}e^{\textsf{i}q\cdot y} \langle 0|\textsf{T} \left\{ \mathcal{J}_{\eta_c}(x)\mathcal{J}_{\Delta,\chi}(y) \bar{\mathcal{J}}_{P_D,\mu}(0) \right\}|0\rangle\, ,
\end{eqnarray}
\begin{eqnarray}\label{PI8}
\Pi_{\mu\chi\zeta,8}(p,q) &=& \textsf{i}^2 \int d^4xd^4y e^{\textsf{i}p\cdot x}e^{\textsf{i}q\cdot y} \langle 0|\textsf{T} \left\{ \mathcal{J}_{J/\psi,\zeta}(x)\mathcal{J}_{\Delta,\chi}(y) \bar{\mathcal{J}}_{P_D,\mu}(0) \right\}|0\rangle\, .
\end{eqnarray}
\begin{eqnarray}\label{PI9}
\Pi_{\mu\nu,9}(p,q) &=& \textsf{i}^2 \int d^4xd^4y e^{\textsf{i}p\cdot x}e^{\textsf{i}q\cdot y} \langle 0|\textsf{T} \left\{ \mathcal{J}_{\eta_c}(x)\mathcal{J}_N(y) \bar{\mathcal{J}}_{P_E,\mu\nu}(0) \right\}|0\rangle\, ,
\end{eqnarray}
\begin{eqnarray}\label{PI10}
\Pi_{\mu\nu\zeta,10}(p,q) &=& \textsf{i}^2 \int d^4xd^4y e^{\textsf{i}p\cdot x}e^{\textsf{i}q\cdot y} \langle 0|\textsf{T} \left\{ \mathcal{J}_{J/\psi,\zeta}(x)\mathcal{J}_N(y) \bar{\mathcal{J}}_{P_E,\mu\nu}(0) \right\}|0\rangle\, ,
\end{eqnarray}
\begin{eqnarray}\label{PI11}
\Pi_{\mu\nu\chi,11}(p,q) &=& \textsf{i}^2 \int d^4xd^4y e^{\textsf{i}p\cdot x}e^{\textsf{i}q\cdot y} \langle 0|\textsf{T} \left\{ \mathcal{J}_{\eta_c}(x)\mathcal{J}_{\Delta,\chi}(y) \bar{\mathcal{J}}_{P_F,\mu\nu}(0) \right\}|0\rangle\, ,
\end{eqnarray}
\begin{eqnarray}\label{PI12}
\Pi_{\mu\nu\chi\zeta,12}(p,q) &=& \textsf{i}^2 \int d^4xd^4y e^{\textsf{i}p\cdot x}e^{\textsf{i}q\cdot y} \langle 0|\textsf{T} \left\{ \mathcal{J}_{J/\psi,\zeta}(x)\mathcal{J}_{\Delta,\chi}(y) \bar{\mathcal{J}}_{P_F,\mu\nu}(0) \right\}|0\rangle\,,
\end{eqnarray}
where $\textsf{T}$ is the time order operator, and $\textsf{i}^2=-1$. On the hadronic side, the complete sets of intermediate hadron states with the same quantum numbers as the currents
$\mathcal{J}_{\eta_c}(x)$, $\mathcal{J}_{J/\psi,\mu}(x)$, $\mathcal{J}_{N}(x)$, $\mathcal{J}_{\Delta,\mu}(x)$ and $\mathcal{J}_{P_{A\sim F}}(x)$ are routinely inserted into those three-point correlation functions, and the contributions of the ground states are isolated, thus, the correlation functions on the hadronic side are given by 
\begin{eqnarray}\label{Correlation1}
\Pi_{\mu,1}(p,q) &=& \frac{f_{\eta_c}m_{\eta_c}^2}{2m_c}\lambda_N\lambda_{P_A} g_{\eta_cN,1} \frac{u(q)\bar{u}(q)\textsf{i}\gamma_5U_{\alpha}(p^{\prime})\bar{U}_{\mu}(p^{\prime})p^{\alpha}}
{\left(m^2_{P_A}-p'^2\right)\left(m^2_{\eta_c}-p^2\right)\left(m^2_{N}-q^2\right)}+\cdots\, ,
\end{eqnarray}

\begin{eqnarray}\label{Correlation2}
\Pi_{\mu\zeta,2}(p,q) &=& f_{J/\psi}m_{J/\psi}\lambda_N\lambda_{P_A}  \frac{-\textsf{i} g_{J/\psi N,2}u(q)\bar{u}(q)U_{\alpha}(p^{\prime})\bar{U}_{\mu}(p^{\prime})\varepsilon_{\zeta}(p)\varepsilon^{*\alpha}(p)}
{\left(m^2_{P_A}-p'^2\right)\left(m^2_{J/\psi}-p^2\right)\left(m^2_{N}-q^2\right)}+\cdots\, ,
\end{eqnarray}

\begin{eqnarray}\label{Correlation3}
\Pi_{\mu\chi,3}(p,q) &=& -\frac{f_{\eta_c}m_{\eta_c}^2}{2m_c}\lambda_{\Delta}\lambda_{P_B} g_{\eta_c\Delta,3} \frac{u_{\chi}(q)\bar{u}_{\beta}(q)U^{\beta}(p^{\prime})\bar{U}_{\mu}(p^{\prime})}
{\left(m^2_{P_B}-p'^2\right)\left(m^2_{\eta_c}-p^2\right)\left(m^2_{\Delta}-q^2\right)}+\cdots\, ,
\end{eqnarray}

\begin{eqnarray}\label{Correlation4}
\Pi_{\mu\chi\zeta,4}(p,q) &=& f_{J/\psi}m_{J/\psi}\lambda_{\Delta}\lambda_{P_B} g_{J/\psi \Delta,4} \nonumber\\ &&\cdot\frac{u_{\chi}(q)\bar{u}_{\beta}(q)\gamma^5\gamma^{\alpha}U_{\beta}(p^{\prime})\bar{U}_{\mu}(p^{\prime})\varepsilon_{\zeta}(p)\varepsilon^{*\alpha}(p)}
{\left(m^2_{P_B}-p'^2\right)\left(m^2_{J/\psi}-p^2\right)\left(m^2_{\Delta}-q^2\right)}+\cdots\, ,
\end{eqnarray}

\begin{eqnarray}\label{Correlation5}
\Pi_{\mu,5}(p,q) &=& \frac{f_{\eta_c}m_{\eta_c}^2}{2m_c}\lambda_N\lambda_{P_C} g_{\eta_cN,5} \frac{u(q)\bar{u}(q)\textsf{i}\gamma^5U_{\alpha}(p^{\prime})\bar{U}_{\mu}(p^{\prime})p^{\alpha}}
{\left(m^2_{P_C}-p'^2\right)\left(m^2_{\eta_c}-p^2\right)\left(m^2_{N}-q^2\right)}+\cdots\, ,
\end{eqnarray}

\begin{eqnarray}\label{Correlation6}
\Pi_{\mu\zeta,6}(p,q) &=&-f_{J/\psi}m_{J/\psi}\lambda_{N}\lambda_{P_C} g_{J/\psi N,6}\frac{u(q)\bar{u}(q)U_{\alpha}(p^{\prime})\bar{U}_{\mu}(p^{\prime})\varepsilon_{\zeta}(p)\varepsilon^{*\alpha}(p)}
{\left(m^2_{P_C}-p'^2\right)\left(m^2_{J/\psi}-p^2\right)\left(m^2_{N}-q^2\right)}+\cdots\, ,
\end{eqnarray}

\begin{eqnarray}\label{Correlation7}
\Pi_{\mu\chi,7}(p,q) &=& \frac{f_{\eta_c}m_{\eta_c}^2}{2m_c}\lambda_\Delta\lambda_{P_D} g_{\eta_c\Delta,7} \frac{\textsf{i} u_{\chi}(q)\bar{u}_{\beta}(q)U^{\beta}(p^{\prime})\bar{U}_{\mu}(p^{\prime})}
{\left(m^2_{P_D}-p'^2\right)\left(m^2_{\eta_c}-p^2\right)\left(m^2_{\Delta}-q^2\right)}+\cdots\, ,
\end{eqnarray}

\begin{eqnarray}\label{Correlation8}
\Pi_{\mu\chi\zeta,8}(p,q) &=&f_{J/\psi}m_{J/\psi}\lambda_{\Delta}\lambda_{P_D} g_{J/\psi \Delta,8}\nonumber\\
&&\cdot\frac{u_{\chi}(q)\bar{u}_{\beta}(q)\gamma^5\gamma_{\alpha}U^{\beta}(p^{\prime})\bar{U}_{\mu}(p^{\prime})\varepsilon_{\zeta}(p)\varepsilon^{*\alpha}(p)}
{\left(m^2_{P_D}-p'^2\right)\left(m^2_{J/\psi}-p^2\right)\left(m^2_{\Delta}-q^2\right)}+\cdots\, ,
\end{eqnarray}

\begin{eqnarray}\label{Correlation9}
\Pi_{\mu\nu,9}(p,q) &=& \frac{\sqrt{2}f_{\eta_c}m_{\eta_c}^2}{2m_c}\lambda_N\lambda_{P_E} g_{\eta_c N,9} \frac{-\textsf{i} u(q)\bar{u}(q)U_{\alpha\beta}(p^{\prime})\bar{U}_{\mu\nu}(p^{\prime})p^{\alpha}p^{\beta}}
{\left(m^2_{P_E}-p'^2\right)\left(m^2_{\eta_c}-p^2\right)\left(m^2_{N}-q^2\right)}+\cdots\, ,
\end{eqnarray}

\begin{eqnarray}\label{Correlation10}
\Pi_{\mu\nu\zeta,10}(p,q) &=&\sqrt{2}f_{J/\psi}m_{J/\psi}\lambda_{N}\lambda_{P_E} g_{J/\psi N,10}\nonumber\\
&&\cdot\frac{u(q)\bar{u}(q)\gamma^5\gamma_{\chi}U_{\alpha\beta}(p^{\prime})\bar{U}_{\mu\nu}(p^{\prime})\varepsilon_{\zeta}(p)\varepsilon^{*\chi}(p)p^{\alpha}p^{\beta}}
{\left(m^2_{P_E}-p'^2\right)\left(m^2_{J/\psi}-p^2\right)\left(m^2_{N}-q^2\right)}+\cdots\, ,
\end{eqnarray}

\begin{eqnarray}\label{Correlation11}
\Pi_{\mu\nu\chi,11}(p,q) &=& \frac{\sqrt{2}f_{\eta_c}m_{\eta_c}^2}{2m_c}\lambda_{\Delta}\lambda_{P_F} g_{\eta_c \Delta,11} \frac{-\textsf{i} u_{\chi}(q)\bar{u}^{\alpha}(q)\gamma^5U_{\alpha\beta}(p^{\prime})\bar{U}_{\mu\nu}(p^{\prime})p^{\beta}}
{\left(m^2_{P_F}-p'^2\right)\left(m^2_{\eta_c}-p^2\right)\left(m^2_{\Delta}-q^2\right)}+\cdots\, ,
\end{eqnarray}

\begin{eqnarray}\label{Correlation12}
\Pi_{\mu\nu\chi\zeta,12}(p,q) &=&\sqrt{2}f_{J/\psi}m_{J/\psi}\lambda_{\Delta}\lambda_{P_F} g_{J/\psi \Delta,12}\nonumber\\
&&\cdot\frac{u_{\chi}(q)\bar{u}^{\xi}(q)\gamma_{\vartheta}U_{\xi\alpha}(p^{\prime})\bar{U}_{\mu\nu}(p^{\prime})\varepsilon_{\zeta}(p)\varepsilon^{*\vartheta}(p)p^{\alpha}}
{\left(m^2_{P_F}-p'^2\right)\left(m^2_{J/\psi}-p^2\right)\left(m^2_{\Delta}-q^2\right)}+\cdots\, ,
\end{eqnarray}
where the  $u(q)$ is the Dirac spinor, the $u_{\mu}(q)$, $U_{\mu}(p^\prime)$ and $U_{\mu\nu}(p^\prime)$  are the Rarita-Schwinger spinors, the $\varepsilon_\zeta$ represents the polarization vector of the $J/\psi$, they follow the equations,
\begin{eqnarray}
\sum_s u(q)\bar{u}(q)&=& \!\not\!{q}+m_N \,,\nonumber \\
\sum_s u_\mu(q)\bar{u}_\nu(q)&=& (\!\not\!{q}+m_\Delta)\left( -g_{\mu\nu}+\frac{\gamma_\mu\gamma_\nu}{3}+\frac{2q_\mu q_\nu}{3q^2}-\frac{q_\mu\gamma_\nu-q_\nu\gamma_\mu}{3\sqrt{q^2}}\right) \,,\nonumber \\
\sum_s U_\mu(p^{\prime})\bar{U}_\nu(p^{\prime})&=& (\!\not\!{p^{\prime}}+m_{P_{A/B/C/D}})\left( -g_{\mu\nu}+\frac{\gamma_\mu\gamma_\nu}{3}+\frac{2p^{\prime}_\mu p^{\prime}_\nu}{3p^{\prime 2}}-\frac{p^{\prime}_\mu\gamma_\nu-p^{\prime}_\nu\gamma_\mu}{3\sqrt{p^{\prime 2}}}\right) \,,
\end{eqnarray}
\begin{eqnarray}
\sum_s U_{\mu\nu}(p^{\prime})\bar{U}_{\alpha\beta}(p^{\prime})&=& (\!\not\!{p^{\prime}}+m_{P_{E/F}})\Bigg\{\frac{\tilde{g}_{\mu\alpha}(p')\tilde{g}_{\nu\beta}(p')+\tilde{g}_{\mu\beta}(p')\tilde{g}_{\nu\alpha}(p')}{2}-\frac{\tilde{g}_{\mu\nu}(p')\tilde{g}_{\alpha\beta}(p')}{5}\nonumber\\
&&-\frac{1}{10}\left(\gamma_\mu\gamma_\alpha+\frac{\gamma_\mu p'_\alpha-\gamma_\alpha p'_\mu}{\sqrt{p'^2}}-\frac{p'_\mu p'_\alpha}{p'^2}\right)\tilde{g}_{\nu\beta}(p')\nonumber\\
&&-\frac{1}{10}\left(\gamma_\nu\gamma_\alpha+\frac{\gamma_\nu p'_\alpha-\gamma_\alpha p'_\nu}{\sqrt{p'^2}}-\frac{p'_\nu p'_\alpha}{p'^2}\right)\tilde{g}_{\mu\beta}(p')\nonumber\\
&&-\frac{1}{10}\left(\gamma_\mu\gamma_\beta+\frac{\gamma_\mu p'_\beta-\gamma_\beta p'_\mu}{\sqrt{p'^2}}-\frac{p'_\mu p'_\beta}{p'^2}\right)\tilde{g}_{\nu\alpha}(p')\nonumber\\
&&-\frac{1}{10}\left(\gamma_\nu\gamma_\beta+\frac{\gamma_\nu p'_\beta-\gamma_\beta p'_\nu}{\sqrt{p'^2}}-\frac{p'_\nu p'_\beta}{p'^2}\right)\tilde{g}_{\mu\alpha}(p')
\Bigg\}
\end{eqnarray}
where, $\tilde{g}_{\mu\nu}(p')=g_{\mu\nu}-\frac{p'_\mu p'_\nu}{p'^2}$, $\varepsilon_{\mu}(p)\varepsilon_{\nu}^*(p)=-\tilde{g}_{\mu\nu}(p)$, $\lambda_N$ and $\lambda_\Delta$ are the pole residues of the $N$ are $\Delta$ baryons, respectively, $\lambda_{P_{A\sim F}}$ represent the pole residues of the states $P_c(4380)$, $P_c(4410)$, $P_c(4440)$, $P_c(4470)$, $P_c(4457)$ and $P_c(4620)$. $f_{\eta_c}$ and $f_{J/\psi}$ are the decay constants of the $\eta_c$ and $J/\psi$ mesons, respectively, $g_{\eta_cN,1/5/9}$ and $g_{J/\psi N,2/6/10}$ denote the strong decay constants for the decay channels $\mathcal{P}_{A/C/E}\rightarrow \eta_cN$ and $\mathcal{P}_{A/C/E}\rightarrow J/\psi N$ of the $P_c(4380)$, $P_c(4440)$ and $P_c(4457)$, respectively, $g_{\eta_c\Delta,3/7/11}$ and $g_{J/\psi\Delta,4/8/12}$ are the strong decay constants for the decay channels $\mathcal{P}_{B/D/F}\rightarrow \eta_c\Delta$ and $\mathcal{P}_{B/D/F}\rightarrow J/\psi\Delta$ of the $P_c(4410)$, $P_c(4470)$ and $P_c(4620)$, respectively, as shown, $\mathcal{P}_{A\sim F}$ represent the above six $P_c$ states. The listed constants are defined as,
\begin{eqnarray}
\langle 0| \mathcal{J}_{N}(0)|N(q)\rangle &=& \lambda_N u(q)\,,\nonumber \\
\langle 0| \mathcal{J}_{\Delta,\mu}(0)|\Delta(q)\rangle &=& \lambda_{\Delta}u_{\mu}(q)\,,\nonumber \\
\langle 0| \mathcal{J}_{J/\psi,\mu}(0)|J/\psi(p)\rangle &=& f_{J/\psi}m_{J/\psi}\varepsilon_{\mu}\,,\nonumber \\
\langle 0| \mathcal{J}_{\eta_c}(0)|\eta_{c}(p)\rangle &=& \frac{f_{\eta_c}m_{\eta_c}^2}{2m_c}\,,
\end{eqnarray}

\begin{eqnarray}
\langle 0| \mathcal{J}_{P_{A/B/C/D},\mu}(0)|\mathcal{P}_{{A/B/C/D}}(p')\rangle &=& \lambda_{P_{{A/B/C/D}}}U_{\mu}(p')\,,\nonumber  \\
\langle 0| \mathcal{J}_{P_{E/F},\mu\nu}(0)|\mathcal{P}_{{E/F}}(p')\rangle &=& \sqrt{2}\lambda_{P_{E/F}}U_{\mu\nu}(p')\,,
\end{eqnarray}

\begin{eqnarray}
\langle \eta_c(p)N(q)| \mathcal{P}_A(p')\rangle &=& \textsf{i} g_{\eta_cN,1} \bar{u}(q)\textsf{i} \gamma_5 U_{\alpha}(p')p^{\alpha}\,,\nonumber  \\
\langle J/\psi(p)N(q)| \mathcal{P}_A(p')\rangle &=& g_{J/\psi N,2} \bar{u}(q) U_{\alpha}(p')\varepsilon^{*\alpha}(p) \,,
\end{eqnarray}
\begin{eqnarray}
\langle \eta_c(p)\Delta(q)| \mathcal{P}_B(p')\rangle &=& -\textsf{i} g_{\eta_c\Delta,3} \bar{u}_\beta(q)U^{\beta}(p')\,,\nonumber  \\
\langle J/\psi(p)\Delta(q)| \mathcal{P}_B(p')\rangle &=& \textsf{i} g_{J/\psi \Delta,4} \bar{u}^\beta(q)\gamma_5\gamma_{\alpha} U_{\beta}(p')\varepsilon^{*\alpha}(p) \,,
\end{eqnarray}
\begin{eqnarray}
\langle \eta_c(p)N(q)| \mathcal{P}_C(p')\rangle &=& \textsf{i} g_{\eta_cN,5} \bar{u}(q)\textsf{i} \gamma_5 U_{\alpha}(p')p^{\alpha}\,,\nonumber  \\
\langle J/\psi(p)N(q)| \mathcal{P}_C(p')\rangle &=& -\textsf{i} g_{J/\psi N,6} \bar{u}(q) U_{\alpha}(p')\varepsilon^{*\alpha}(p) \,,
\end{eqnarray}
\begin{eqnarray}
\langle \eta_c(p)\Delta(q)| \mathcal{P}_D(p')\rangle &=& -\textsf{i} g_{\eta_c\Delta,7} \bar{u}_\beta(q)U^{\beta}(p')\,,\nonumber  \\
\langle J/\psi(p)\Delta(q)| \mathcal{P}_D(p')\rangle &=& \textsf{i} g_{J/\psi \Delta,8} \bar{u}^\beta(q)\gamma_5\gamma_{\alpha} U_{\beta}(p')\varepsilon^{*\alpha}(p) \,,
\end{eqnarray}
\begin{eqnarray}
\langle \eta_c(p)N(q)| \mathcal{P}_E(p')\rangle &=& g_{\eta_cN,9} \bar{u}(q)U_{\alpha\beta}(p')p^{\alpha}p^{\beta}\,,\nonumber  \\
\langle J/\psi(p)N(q)| \mathcal{P}_E(p')\rangle &=& g_{J/\psi N,10} \bar{u}(q)\textsf{i} \gamma_5\gamma_{\chi} U_{\alpha\beta}(p')p^{\alpha}p^{\beta}\varepsilon^{*\chi}(p) \,,
\end{eqnarray}
\begin{eqnarray}
\langle \eta_c(p)\Delta(q)| \mathcal{P}_F(p')\rangle &=& -\textsf{i} g_{\eta_c\Delta,11} \bar{u}^\chi(q)\textsf{i} \gamma_5 U_{\chi\beta}(p')p^{\beta}\,,\nonumber  \\
\langle J/\psi(p)\Delta(q)| \mathcal{P}_F(p')\rangle &=& \textsf{i} g_{J/\psi \Delta,12} \bar{u}^\xi(q)\gamma_{\vartheta} U_{\xi\chi}(p')p^{\chi}\varepsilon^{*\vartheta}(p) \,,
\end{eqnarray}
where, $m_c$, $m_{\eta_c}$, $m_{J/\psi}$, $m_N$, $m_{\Delta}$ and $m_{P_{A\sim F}}$ are the masses of the charm quark, $\eta_c$, $J/\psi$, $N$, $\Delta$ and $\mathcal{P}_{P_{A\sim F}}$, respectively, $|\eta_c\rangle$, $|J/\psi\rangle$, $|N\rangle$, $|\Delta\rangle$, and $|\mathcal{P}_{A\sim F}\rangle$ denote the ground states of $\eta_c$, $J/\psi$, $N$, $\Delta$ and $\mathcal{P}_{A\sim F}$, respectively.

The correlation functions at both the hadronic and QCD sides are the matrices with complicated structures in the Dirac spinor space, it is reasonable to consider that the two sides should match each other, namely, $\Pi_{H}(p,q)=\Pi_{QCD}(p,q)$. In the framework of the QCD sum rules, the decay constants should not depend on the detailed structures, for example, the same decay constants are derived via different structures in Ref. \cite{Myself-DcPc4312}. The equations $Tr[\Pi_{H}(p,q)\cdot\Gamma]=Tr[\Pi_{QCD}(p,q)\cdot\Gamma]$ are applied to select the detailed structures, where $\Gamma$ is some chosen $\gamma-$matrix in the Dirac spinor space. The selected structures for both the hadronic and QCD sides are expressed as follows:
\begin{eqnarray}
\Pi_{\mu,1}(p,q)\cdot \textsf{i}\gamma_5\tilde{g}_{\mu\nu}(p^\prime) &=& \Pi_1(p'^2,p^2,q^2)\!\not\!{p}\!\not\!{q}q_{\nu}+\cdots\,,\nonumber \\
\Pi_{\mu\zeta,2}(p,q)\tilde{g}_{\mu\nu}(p^\prime) &=& \Pi_2(p'^2,p^2,q^2)\!\not\!{p}\!\not\!{q}q_{\nu}q_{\zeta}+\cdots\,,\nonumber \\
\Pi_{\mu\chi,3}(p,q)\tilde{g}_{\mu\nu}(p^\prime) &=& \Pi_3(p'^2,p^2,q^2)\!\not\!{p}\!\not\!{q}q_{\nu}q_{\chi}+\cdots\,,\nonumber \\
\Pi_{\mu\chi\zeta,4}(p,q)\tilde{g}_{\mu\nu}(p^\prime) &=& \Pi_4(p'^2,p^2,q^2)\!\not\!{p}\!\not\!{q}q_{\nu}g_{\zeta\chi}+\cdots\,,\nonumber \\
\Pi_{\mu,5}(p,q)\cdot \textsf{i}\gamma_5\tilde{g}_{\mu\nu}(p^\prime) &=& \Pi_5(p'^2,p^2,q^2)\!\not\!{p}\!\not\!{q}q_{\nu}+\cdots\,,\nonumber \\
\Pi_{\mu\zeta,6}(p,q)\tilde{g}_{\mu\nu}(p^\prime) &=& \Pi_6(p'^2,p^2,q^2)\!\not\!{p}\!\not\!{q}q_{\nu}q_{\zeta}+\cdots\,,\nonumber \\
\Pi_{\mu\chi,7}(p,q) \tilde{g}_{\mu\nu}(p^\prime) &=& \Pi_7(p'^2,p^2,q^2)\!\not\!{p}\!\not\!{q}q_{\nu}q_{\chi}+\cdots\,,\nonumber \\
\Pi_{\mu\chi\zeta,8}(p,q)\tilde{g}_{\mu\nu}(p^\prime) &=& \Pi_8(p'^2,p^2,q^2)\!\not\!{p}\!\not\!{q}q_{\nu}g_{\zeta\chi}+\cdots\,,\nonumber \\
\Pi_{\mu\nu,9}(p,q) \tilde{g}_{\mu\rho}(p^\prime)\tilde{g}_{\nu\sigma}(p^\prime) &=& \Pi_9(p'^2,p^2,q^2)\!\not\!{p}\!\not\!{q}g_{\rho\sigma}+\cdots\,,\nonumber \\
\Pi_{\mu\nu\zeta,10}(p,q)\cdot \gamma_5\tilde{g}_{\mu\rho}(p^\prime)\tilde{g}_{\nu\sigma}(p^\prime) &=& \Pi_{10}(p'^2,p^2,q^2)\!\not\!{p}\!\not\!{q}q_{\zeta}g_{\rho\sigma}+\cdots\,,\nonumber \\
\Pi_{\mu\nu\chi,11}(p,q)\cdot \textsf{i} \gamma_5 \tilde{g}_{\mu\rho}(p^\prime)\tilde{g}_{\nu\sigma}(p^\prime) &=& \Pi_{11}(p'^2,p^2,q^2)\!\not\!{p}\!\not\!{q}q_{\chi}g_{\rho\sigma}+\cdots\,,\nonumber \\
\Pi_{\mu\nu\chi\zeta,12}(p,q)\tilde{g}_{\mu\rho}(p^\prime)\tilde{g}_{\nu\sigma}(p^\prime) &=& \Pi_{12}(p'^2,p^2,q^2)\!\not\!{p}\!\not\!{q}q_{\chi}q_{\zeta}g_{\rho\sigma}+\cdots\,.
\end{eqnarray}
For the $J^P=\frac{5}{2}^-$, the two-point correlation function determined by the current $\mathcal{J}_{P_E,\mu\nu}(x)$ or $\mathcal{J}_{P_F,\mu\nu}(x)$ on the hadron side can be written as \cite{ZGWang-Review4},
\begin{eqnarray}
\Pi_{\mu\nu\alpha\beta}(p^\prime)&=&2{\lambda_{\frac{5}{2}}^-}^2\frac{\!\not\!{p^\prime+m_{-}}}{m_{-}^2-{\!\not\!p^\prime}^2}\Bigg[\frac{\tilde{g}_{\mu\alpha}\tilde{g}_{\nu\beta}+\tilde{g}_{\mu\beta}\tilde{g}_{\nu\alpha}}{2}
-\frac{\tilde{g}_{\mu\nu}\tilde{g}_{\alpha\beta}}{5}-\frac{1}{10}\left(\gamma_\mu\gamma_\alpha+\frac{\gamma_\mu p^\prime_\alpha-\gamma_\alpha p^\prime_\mu}{\sqrt{{p^\prime}^2}}-\frac{p^\prime_\mu p^\prime_\alpha}{{p^\prime}^2}\right)\nonumber\\
&& \cdot\tilde{g}_{\nu\beta}-\frac{1}{10}\left(\gamma_\nu\gamma_\alpha+\frac{\gamma_\nu p^\prime_\alpha-\gamma_\alpha p^\prime_\nu}{\sqrt{{p^\prime}^2}}-\frac{p^\prime_\nu p^\prime_\alpha}{{p^\prime}^2}\right)\tilde{g}_{\mu\beta}+\cdot\cdot\cdot\Bigg] \nonumber\\
&& +{f_{\frac{3}{2}}^{-}}^2\frac{\!\not\!{p^\prime-m_{-}}}{m_{-}^2-{\!\not\!p^\prime}^2}\Bigg[p^\prime_\mu p^\prime_\alpha\left(-g_{\nu\beta}+\frac{\gamma_\nu\gamma_\beta}{3}+\frac{2p^\prime_\nu p^\prime_\beta}{3{p^\prime}^2}-\frac{p^\prime_\nu\gamma_\beta-p^\prime_\beta\gamma_\nu}{3\sqrt{{p^\prime}^2}}\right)+\cdot\cdot\cdot\Bigg]\nonumber\\
&& +{h_{\frac{1}{2}}^-}^2\frac{\!\not\!{p^\prime+m_{-}}}{m_{-}^2-{\!\not\!p^\prime}^2}p^\prime_\mu p^\prime_\nu p^\prime_\alpha p^\prime_\beta+\cdot\cdot\cdot\,,
\end{eqnarray}
where, we have already selected the negative parity, the items containing the parameters $f_{\frac{3}{2}}^{-}$ and $h_{\frac{1}{2}}^-$ are the components coupling to angular momentum $\frac{3}{2}$ and $\frac{1}{2}$, respectively. Obviously,  $\mathcal{F}_{\frac{5}{2}}p^\prime_\mu p^\prime_\alpha=0$ and $\mathcal{F}_{\frac{5}{2}}p^\prime_\mu p^\prime_\nu p^\prime_\alpha p^\prime_\beta=0$, where, $\mathcal{F}_{\frac{5}{2}}=\tilde{g}_{\mu\nu}(p^{\prime})\tilde{g}_{\alpha\beta}(p^{\prime})$ is the projector used to obtain the component of the correlation function with the angular momentum $J=\frac{5}{2}$. One could easily find that the projector for the $J=\frac{3}{2}$ can be set as $\mathcal{F}_{\frac{3}{2}}=\tilde{g}_{\mu\nu}(p^{\prime})$.

On the QCD sides, Wick theorem is applied, the correlation functions are then expressed in terms of the full propagators. Followed by the operator product expansion, the traces are performed and the selected tensor structures are chosen, thus, $\Pi_{1\sim 12}(p'^2,p^2,q^2)$ are derived on the QCD sides. Since $p'=p+q$ for the strong decays, considering the expressions of the correlation functions for Eqs.(25-36) and taking the Eq.(25) as an example, it is impossible to deal with the correlation function strictly with the denominator $\left[m^2_{P_A}-(p+q)^2\right]\left(m^2_{\eta_c}-p^2\right)\left(m^2_{N}-q^2\right)$. Similar situation occur for the correlation function on the QCD side. $p'^2$ is approximately set as $\xi p^2$, where the $\xi$ is a parameter. One can show that $0\leq \xi\leq \frac{2q^2}{p^2}+2$. For the decay $P_c(4380)\rightarrow\eta_c+N$, $0\leq \xi\leq \frac{2m_N^2}{m_{\eta_c}^2}+2$, $\xi$ is set as $\frac{m_N^2}{m_{\eta_c^2}}+1$. Following our previous works \cite{Decay-mole-WZG-WX,WZG-ZJX-Zc-Decay,WZG-Y4660-Decay,WZG-X4140-decay,WZG-X4274-decay,WZG-Z4600-decay,WZG-Pc4312-decay-tetra}, rigorous quark-hadron duality below the continuum thresholds is taken, and the double Borel transformation is performed. The QCD sum rules for the hadronic coupling constants are given by,
\begin{eqnarray}\label{QCDG1}
&& \frac{f_{\eta_c}m_{\eta_c}^2\lambda_N\lambda_{P_A} g_{\eta_cN,1}}{2m_c \xi}\frac{\kappa_1}{\frac{m_{P_A}^2}{\xi}-m_{\eta_c}^2}\left\{ {\rm exp} \left( -\frac{m_{\eta_c}^2}{T_1^2} \right)-{\rm exp} \left( -\frac{m_{P_A}^2}{\xi T_1^2} \right) \right\}{\rm exp} \left( -\frac{m_N^2}{T_2^2} \right) \nonumber \\
&& +\mathcal{C}_1 {\rm exp} \left( -\frac{m_{\eta_c}^2}{T_1^2}-\frac{m_N^2}{T_2^2} \right)=\int_{4m_c^2}^{s_{\eta_c}^0}ds\int_0^{s_N^0}du\, \rho_1(s,u){\rm exp}\left( -\frac{s}{T_1^2}-\frac{u}{T_2^2} \right)\,,
\end{eqnarray}

\begin{eqnarray}\label{QCDG2}
&& \frac{f_{J/\psi}m_{J/\psi}\lambda_{N}\lambda_{P_A} g_{J/\psi N,2}}{\xi}\frac{\kappa_2}{\frac{m_{P_A}^2}{\xi}-m_{J/\psi}^2}\left\{ {\rm exp} \left( -\frac{m_{J/\psi}^2}{T_1^2} \right)-{\rm exp} \left( -\frac{m_{P_A}^2}{\xi T_1^2} \right) \right\}{\rm exp} \left( -\frac{m_N^2}{T_2^2} \right) \nonumber \\
&& +\mathcal{C}_2 {\rm exp} \left( -\frac{m_{J/\psi}^2}{T_1^2}-\frac{m_N^2}{T_2^2} \right)=\int_{4m_c^2}^{s_{J/\psi}^0}ds\int_0^{s_N^0}du\, \rho_2(s,u){\rm exp}\left( -\frac{s}{T_1^2}-\frac{u}{T_2^2} \right)\,,
\end{eqnarray}

\begin{eqnarray}\label{QCDG3}
&& \frac{f_{\eta_c}m_{\eta_c}^2\lambda_\Delta\lambda_{P_B} g_{\eta_c\Delta,3}}{2m_c \xi}\frac{\kappa_3}{\frac{m_{P_B}^2}{\xi}-m_{\eta_c}^2}\left\{ {\rm exp} \left( -\frac{m_{\eta_c}^2}{T_1^2} \right)-{\rm exp} \left( -\frac{m_{P_B}^2}{\xi T_1^2} \right) \right\}{\rm exp} \left( -\frac{m_\Delta^2}{T_2^2} \right) \nonumber \\
&& +\mathcal{C}_3 {\rm exp} \left( -\frac{m_{\eta_c}^2}{T_1^2}-\frac{m_\Delta^2}{T_2^2} \right)=\int_{4m_c^2}^{s_{\eta_c}^0}ds\int_0^{s_\Delta^0}du\, \rho_3(s,u){\rm exp}\left( -\frac{s}{T_1^2}-\frac{u}{T_2^2} \right)\,,
\end{eqnarray}

\begin{eqnarray}\label{QCDG4}
&& \frac{f_{J/\psi}m_{J/\psi}\lambda_{\Delta}\lambda_{P_B} g_{J/\psi \Delta,4}}{\xi}\frac{\kappa_4}{\frac{m_{P_B}^2}{\xi}-m_{J/\psi}^2}\left\{ {\rm exp} \left( -\frac{m_{J/\psi}^2}{T_1^2} \right)-{\rm exp} \left( -\frac{m_{P_B}^2}{\xi T_1^2} \right) \right\}{\rm exp} \left( -\frac{m_\Delta^2}{T_2^2} \right) \nonumber \\
&& +\mathcal{C}_4 {\rm exp} \left( -\frac{m_{J/\psi}^2}{T_1^2}-\frac{m_\Delta^2}{T_2^2} \right)=\int_{4m_c^2}^{s_{J/\psi}^0}ds\int_0^{s_\Delta^0}du\, \rho_4(s,u){\rm exp}\left( -\frac{s}{T_1^2}-\frac{u}{T_2^2} \right)\,,
\end{eqnarray}

\begin{eqnarray}\label{QCDG5}
&& \frac{f_{\eta_c}m_{\eta_c}^2\lambda_N\lambda_{P_C} g_{\eta_cN,5}}{2m_c \xi}\frac{\kappa_5}{\frac{m_{P_C}^2}{\xi}-m_{\eta_c}^2}\left\{ {\rm exp} \left( -\frac{m_{\eta_c}^2}{T_1^2} \right)-{\rm exp} \left( -\frac{m_{P_C}^2}{\xi T_1^2} \right) \right\}{\rm exp} \left( -\frac{m_N^2}{T_2^2} \right) \nonumber \\
&& +\mathcal{C}_5 {\rm exp} \left( -\frac{m_{\eta_c}^2}{T_1^2}-\frac{m_N^2}{T_2^2} \right)=\int_{4m_c^2}^{s_{\eta_c}^0}ds\int_0^{s_N^0}du\, \rho_5(s,u){\rm exp}\left( -\frac{s}{T_1^2}-\frac{u}{T_2^2} \right)\,,
\end{eqnarray}

\begin{eqnarray}\label{QCDG6}
&& \frac{f_{J/\psi}m_{J/\psi}\lambda_{N}\lambda_{P_C} g_{J/\psi N,6}}{\xi}\frac{\kappa_6}{\frac{m_{P_C}^2}{\xi}-m_{J/\psi}^2}\left\{ {\rm exp} \left( -\frac{m_{J/\psi}^2}{T_1^2} \right)-{\rm exp} \left( -\frac{m_{P_C}^2}{\xi T_1^2} \right) \right\}{\rm exp} \left( -\frac{m_N^2}{T_2^2} \right) \nonumber \\
&& +\mathcal{C}_6 {\rm exp} \left( -\frac{m_{J/\psi}^2}{T_1^2}-\frac{m_N^2}{T_2^2} \right)=\int_{4m_c^2}^{s_{J/\psi}^0}ds\int_0^{s_N^0}du\, \rho_6(s,u){\rm exp}\left( -\frac{s}{T_1^2}-\frac{u}{T_2^2} \right)\,,
\end{eqnarray}

\begin{eqnarray}\label{QCDG7}
&& \frac{f_{\eta_c}m_{\eta_c}^2\lambda_\Delta\lambda_{P_D} g_{\eta_c\Delta,7}}{2m_c \xi}\frac{\kappa_7}{\frac{m_{P_D}^2}{\xi}-m_{\eta_c}^2}\left\{ {\rm exp} \left( -\frac{m_{\eta_c}^2}{T_1^2} \right)-{\rm exp} \left( -\frac{m_{P_D}^2}{\xi T_1^2} \right) \right\}{\rm exp} \left( -\frac{m_\Delta^2}{T_2^2} \right) \nonumber \\
&& +\mathcal{C}_7 {\rm exp} \left( -\frac{m_{\eta_c}^2}{T_1^2}-\frac{m_\Delta^2}{T_2^2} \right)=\int_{4m_c^2}^{s_{\eta_c}^0}ds\int_0^{s_\Delta^0}du\, \rho_7(s,u){\rm exp}\left( -\frac{s}{T_1^2}-\frac{u}{T_2^2} \right)\,,
\end{eqnarray}

\begin{eqnarray}\label{QCDG8}
&& \frac{f_{J/\psi}m_{J/\psi}\lambda_{\Delta}\lambda_{P_D} g_{J/\psi \Delta,8}}{\xi}\frac{\kappa_8}{\frac{m_{P_D}^2}{\xi}-m_{J/\psi}^2}\left\{ {\rm exp} \left( -\frac{m_{J/\psi}^2}{T_1^2} \right)-{\rm exp} \left( -\frac{m_{P_D}^2}{\xi T_1^2} \right) \right\}{\rm exp} \left( -\frac{m_\Delta^2}{T_2^2} \right) \nonumber \\
&& +\mathcal{C}_8 {\rm exp} \left( -\frac{m_{J/\psi}^2}{T_1^2}-\frac{m_\Delta^2}{T_2^2} \right)=\int_{4m_c^2}^{s_{J/\psi}^0}ds\int_0^{s_\Delta^0}du\, \rho_8(s,u){\rm exp}\left( -\frac{s}{T_1^2}-\frac{u}{T_2^2} \right)\,,
\end{eqnarray}

\begin{eqnarray}\label{QCDG9}
&& \frac{f_{\eta_c}m_{\eta_c}^2\lambda_N\lambda_{P_E} g_{\eta_cN,9}}{2m_c \xi}\frac{\sqrt{2}\kappa_9}{\frac{m_{P_E}^2}{\xi}-m_{\eta_c}^2}\left\{ {\rm exp} \left( -\frac{m_{\eta_c}^2}{T_1^2} \right)-{\rm exp} \left( -\frac{m_{P_E}^2}{\xi T_1^2} \right) \right\}{\rm exp} \left( -\frac{m_N^2}{T_2^2} \right) \nonumber \\
&& +\mathcal{C}_9 {\rm exp} \left( -\frac{m_{\eta_c}^2}{T_1^2}-\frac{m_N^2}{T_2^2} \right)=\int_{4m_c^2}^{s_{\eta_c}^0}ds\int_0^{s_N^0}du\, \rho_9(s,u){\rm exp}\left( -\frac{s}{T_1^2}-\frac{u}{T_2^2} \right)\,,
\end{eqnarray}

\begin{eqnarray}\label{QCDG10}
&& \frac{f_{J/\psi}m_{J/\psi}\lambda_{N}\lambda_{P_E} g_{J/\psi N,10}}{\xi}\frac{\sqrt{2}\kappa_{10}}{\frac{m_{P_E}^2}{\xi}-m_{J/\psi}^2}\left\{ {\rm exp} \left( -\frac{m_{J/\psi}^2}{T_1^2} \right)-{\rm exp} \left( -\frac{m_{P_E}^2}{\xi T_1^2} \right) \right\}{\rm exp} \left( -\frac{m_N^2}{T_2^2} \right) \nonumber \\
&& +\mathcal{C}_{10} {\rm exp} \left( -\frac{m_{J/\psi}^2}{T_1^2}-\frac{m_N^2}{T_2^2} \right)=\int_{4m_c^2}^{s_{J/\psi}^0}ds\int_0^{s_N^0}du\, \rho_{10}(s,u){\rm exp}\left( -\frac{s}{T_1^2}-\frac{u}{T_2^2} \right)\,,
\end{eqnarray}

\begin{eqnarray}\label{QCDG11}
&& \frac{f_{\eta_c}m_{\eta_c}^2\lambda_\Delta\lambda_{P_F} g_{\eta_c\Delta,11}}{2m_c \xi}\frac{\sqrt{2}\kappa_{11}}{\frac{m_{P_F}^2}{\xi}-m_{\eta_c}^2}\left\{ {\rm exp} \left( -\frac{m_{\eta_c}^2}{T_1^2} \right)-{\rm exp} \left( -\frac{m_{P_F}^2}{\xi T_1^2} \right) \right\}{\rm exp} \left( -\frac{m_\Delta^2}{T_2^2} \right) \nonumber \\
&& +\mathcal{C}_{11} {\rm exp} \left( -\frac{m_{\eta_c}^2}{T_1^2}-\frac{m_\Delta^2}{T_2^2} \right)=\int_{4m_c^2}^{s_{\eta_c}^0}ds\int_0^{s_\Delta^0}du\, \rho_{11}(s,u){\rm exp}\left( -\frac{s}{T_1^2}-\frac{u}{T_2^2} \right)\,,
\end{eqnarray}

\begin{eqnarray}\label{QCDG12}
&& \frac{f_{J/\psi}m_{J/\psi}\lambda_{\Delta}\lambda_{P_F} g_{J/\psi \Delta,12}}{\xi}\frac{\sqrt{2}\kappa_{12}}{\frac{m_{P_F}^2}{\xi}-m_{J/\psi}^2}\left\{ {\rm exp} \left( -\frac{m_{J/\psi}^2}{T_1^2} \right)-{\rm exp} \left( -\frac{m_{P_F}^2}{\xi T_1^2} \right) \right\}{\rm exp} \left( -\frac{m_\Delta^2}{T_2^2} \right) \nonumber \\
&& +\mathcal{C}_{12} {\rm exp} \left( -\frac{m_{J/\psi}^2}{T_1^2}-\frac{m_\Delta^2}{T_2^2} \right)=\int_{4m_c^2}^{s_{J/\psi}^0}ds\int_0^{s_\Delta^0}du\, \rho_{12}(s,u){\rm exp}\left( -\frac{s}{T_1^2}-\frac{u}{T_2^2} \right)\,,
\end{eqnarray}
where,
\begin{eqnarray}
\kappa_1&=&\frac{m_N m_{P_{A}} [4 (m_{\eta_c}^2 \tau-m_N^2)+8m_{\eta_c}^2]}{3  m_{\eta_c}^2 \xi}-\frac{  (m_{\eta_c}^2 \tau+m_N^2) [2(m_{\eta_c}^2 \tau-m_N^2)+4m_{\eta_c}^2]}{3 m_{\eta_c}^2 \xi}
\end{eqnarray}
and
 \begin{eqnarray}
\rho_{Z}(s,u)&=& {\lim_{\epsilon_2\to 0}} \,\,{\lim_{\epsilon_1\to 0}}\,\,\frac{  {\rm Im}_{s}\,{\rm Im}_{u}\,\Pi_{Z}(p^{\prime 2},s+\textsf{i}\epsilon_2,u+\textsf{i}\epsilon_1) }{\pi^2} \, .
\end{eqnarray}
  The spectral densities at the QCD sides, $\Pi_{Z}$ are the correlation functions at QCD sides selected from the corresponding structures in the same manner as the hadronic sides, where $Z=1\sim12$. The parameters $\kappa_{2\thicksim12}$ are listed in the Appendix, while the complex expressions of the spectral densities at the QCD sides are omitted. The $\mathcal{C}_{1\sim12}$ denote the unknown parameters which are determined in the numerical calculations to obtain the flat platforms \cite{Decay-mole-WZG-WX,WZG-ZJX-Zc-Decay,WZG-Y4660-Decay,WZG-X4140-decay,
  WZG-X4274-decay,WZG-Z4600-decay,WZG-Pc4312-decay-tetra}. Now, we take the decay channel $P_c(4380)\rightarrow\eta_c+N$ as an example to make a brief interpretation for origin of these free parameters. Taking the triple dispersion relation, the correlation function $\Pi_{\mu,1}$ at the hadronic side is given by,
\begin{eqnarray}\label{dispersion-1}
\Pi_{\mu,1,H}(p^{\prime2},p^2,q^2)&=&\int_{\Delta_s^{\prime2}}^\infty ds^{\prime} \int_{\Delta_s^2}^\infty ds \int_{\Delta_u^2}^\infty du \frac{\rho_{H}(s^\prime,s,u)}{(s^\prime-p^{\prime2})(s-p^2)(u-q^2)}\, ,
\end{eqnarray}
where $\Delta_{s}^{\prime2}$, $\Delta_{s}^{2}$ and $\Delta_{u}^{2}$ are thresholds, we introduce the subscript $H$ to denote the hadron side. At the QCD side, the double dispersion relation is applied to acquire,
\begin{eqnarray}\label{dispersion-2}
\Pi_{QCD}(p^{\prime2},p^2,q^2)&=& \int_{\Delta_s^2}^\infty ds \int_{\Delta_u^2}^\infty du \frac{\rho_{QCD}(p^{\prime2},s,u)}{(s-p^2)(u-q^2)}\, ,
\end{eqnarray}
as
\begin{eqnarray}
{\rm lim}_{\epsilon \to 0}\frac{{\rm Im}\,\Pi_{QCD}(s^\prime+i\epsilon,p^2,q^2)}{\pi}&=&0\, .
\end{eqnarray}
Obviously, the triple dispersion relation on the hadron side cannot match with the double dispersion relation on  the QCD side, the integral over $ds^\prime$ is carried out firstly and then match the hadron side with the QCD side below the continuum thresholds to acquire rigorous quark-hadron duality \cite{WZG-ZJX-Zc-Decay,WZG-Y4660-Decay},
\begin{eqnarray}\label{rigorous}
  \int_{\Delta_s^2}^{s_{0}}ds \int_{\Delta_u^2}^{u_0}du  \frac{\rho_{QCD}(p^{\prime2},s,u)}{(s-p^2)(u-q^2)}&=& \int_{\Delta_s^2}^{s_0}ds \int_{\Delta_u^2}^{u_0}du  \left[ \int_{\Delta_{s}^{\prime2}}^{\infty}ds^\prime  \frac{\rho_H(s^\prime,s,u)}{(s^\prime-p^{\prime2})(s-p^2)(u-q^2)} \right]\, ,
\end{eqnarray}
where the $s_0$ and $u_0$ are the continuum thresholds. Then, the free parameter $\mathcal{C}_1$ is introduced to parameterize the contributions of transitions between the  higher resonances (continuum states) in the $s^\prime$ channel and the ground state conventional meson-baryon pair. It is written as,
\begin{eqnarray}
\mathcal{C}_{1}&=&\int_{s_0^\prime}^{\infty} ds^\prime \frac{\rho_H(s^\prime,m_{\eta_c}^2,m_N^2)}{\left(s^\prime-m_{P_A}^2\right)
\left(p^2-m_{\eta_c}^2 \right)\left(q^2-m_N^2 \right)}\, ,
\end{eqnarray}
where the $s_0^\prime$ is the continuum threshold parameter for the ground state,
the $\rho_H(s^\prime,m_{\eta_c}^2,m_N^2)$ is the formal  hadronic spectral density for transitions between the  higher resonances (continuum states) in the $s^\prime$ channel and ground state of the meson-baryon pair $\eta_cN$. The fact is that, for the the hadron side and QCD side of the spectral density below the continuum thresholds $s_0$ and $u_0$ in the $s$ and $u$ channels, they have one to one correspondence, while in the $s^\prime$ channel, there is no corresponding counterpart on the QCD side.
Experimentally, the spectroscopy of the hidden-charm pentaquark states has not been established yet, making it impossible to determine its explicit expression right now. It is reasonable to introduce a free parameter $\mathcal{C}_{1}$ to parameterize the contributions involving the higher resonances (continuum states) in the $s^\prime$ channel, which results in model dependence. At present, we have no choice to avoid model dependence, this is also the reason for us to set its uncertainty $\delta\mathcal{C}_{1}$ to zero. The correctness or not depends on the further experimental progress. For more detailed discussions, one can consult Sect.{\bf 7} in Ref.\cite{ZGWang-Review4}.

\section{Numerical results and discussions}
Based on the detailed expressions of $\rho_{1\sim12}$, the numerical calculation is conducted. As for the vacuum condensates on the QCD sides, the standard values are applied which are listed as $\langle\overline{q}q\rangle=-(0.24\pm0.01\,{\rm GeV})^3$, $\langle\overline{q}g_s\sigma Gq\rangle=m_0^2\langle\overline{q}q\rangle$, $m_0^2=(0.8\pm0.1)\,{\rm GeV}^2$, $\langle\frac{\alpha_s}{\pi}GG\rangle=(0.33\,{\rm GeV})^4$ at the energy scale $\mu=1\,{\rm GeV}$ \cite{SVZ1,SVZ2,Reinders,ColangeloReview}, and the value of the $
\overline{MS}$ mass $m_c(m_c)=1.275\pm0.025\,{\rm GeV}$ is applied from the Particle Data Group \cite{PDG}. The energy-scale dependence of these parameters are written as,
\begin{eqnarray}
\notag \langle\overline{q}q\rangle(\mu)&&=\langle\overline{q}q\rangle(1{\rm GeV})\left[\frac{\alpha_s(1{\rm GeV})}{\alpha_s(\mu)}\right]^{\frac{12}{33-2n_f}}\, ,\\
\notag \langle\overline{q}g_s\sigma Gq\rangle(\mu)&& =\langle\overline{q}g_s\sigma Gq\rangle(1{\rm GeV})\left[\frac{\alpha_s(1{\rm GeV})}{\alpha_s(\mu)}\right]^{\frac{2}{33-2n_f}}\, ,\\
\notag  m_c(\mu)&&=m_c(m_c)\left[\frac{\alpha_s(\mu)}{\alpha_s(m_c)}\right]^{\frac{12}{33-2n_f}}\, ,\\
\notag \alpha_s(\mu)&&=\frac{1}{b_0t}\left[1-\frac{b_1}{b_0^2}\frac{\rm{log}\emph{t}}{t}+\frac{b_1^2(\rm{log}^2\emph{t}-\rm{log}\emph {t}-1)+\emph{b}_0\emph{b}_2}{b_0^4t^2}\right]\, ,
\end{eqnarray}
where $t={\rm log}\frac{\mu^2}{\Lambda_{QCD}^2}$, $\emph b_0=\frac{33-2\emph{n}_\emph{f}}{12\pi}$, $b_1=\frac{153-19n_f}{24\pi^2}$, $b_2=\frac{2857-\frac{5033}{9}n_f+\frac{325}{27}n_f^2}{128\pi^3}$
and $\Lambda_{QCD}=213$ MeV, $296$ MeV, $339$ MeV for the flavors $n_f=5,4,3$, respectively \cite{PDG,Narison}, and $n_f=4$ for the strong decays of the present study. For the decay products containing $\eta_c(J/\psi)$, the energy scale $\mu=\frac{1}{2}m_{\eta_c}(\frac{1}{2}m_{J/\psi})$, respectively \cite{Decay-mole-WZG-WX,EWZGHuang}.

The masses of the states $\mathcal{P}_A$, $\mathcal{P}_C$ and $\mathcal{P}_E$ obey the experimental data, set $m_{P_A}=4.380\,{\rm GeV}$, $m_{P_C}=4.410\,{\rm GeV}$ and $m_{P_E}=4.457\,{\rm GeV}$ \cite{RAaij2}. The masses of the states $\mathcal{P}_B$, $\mathcal{P}_D$ and $\mathcal{P}_F$ follow our conclusion in Ref. \cite{mass-mole-WXW-SCPMA}, and set $m_{P_B}=4.410\,{\rm GeV}$, $m_{P_D}=4.470\,{\rm GeV}$ and $m_{P_F}=4.620\,{\rm GeV}$. The masses of $\eta_c$, $J/\psi$, $N$ and $\Delta$ are from the Particle Data Group \cite{PDG}, taking $m_{\eta_c}=2.984\,{\rm GeV}$, $m_{J/\psi}=3.097\,{\rm GeV}$, $m_N=0.938\,{\rm GeV}$, $m_{\Delta}=1.232\,{\rm GeV}$. The values of the decay constants of the $\eta_c$ and $J/\psi$ follow the results in Ref. \cite{Becirevic}, choosing $f_{J/\psi}=0.418\,{\rm GeV}$, $f_{\eta_c}=0.387\,{\rm GeV}$. For the pole residues, $\lambda_N=3.20\times 10^{-2}\,{\rm GeV^3}$ \cite{EIoffe}, $\lambda_{\Delta}=7.63\times 10^{-3}\,{\rm GeV^3}$ \cite{Myself-DcPc4312}, $\lambda_{P_A}=1.97\times10^{-3}\,{\rm GeV^6}$, $\lambda_{P_B}=1.24\times10^{-3}\,{\rm GeV^6}$, $\lambda_{P_C}=3.60\times10^{-3}\,{\rm GeV^6}$, $\lambda_{P_D}=2.31\times10^{-3}\,{\rm GeV^6}$, $\lambda_{P_E}=4.05\times10^{-3}\,{\rm GeV^6}$, $\lambda_{P_F}=2.40\times10^{-3}\,{\rm GeV^6}$ \cite{mass-mole-WXW-SCPMA}.
For the threshold parameters, they are set as, $\sqrt{s_{\eta_c}^0}=3.50\,{\rm GeV}$, $\sqrt{s_{N}^0}=1.30\,{\rm GeV}$, $\sqrt{s_{J/\psi}^0}=3.60\,{\rm GeV}$ \cite{Decay-mole-WZG-WX} and $\sqrt{s_{\Delta}^0}=1.61\,{\rm GeV}$ \cite{Myself-DcPc4312}.

\renewcommand{\arraystretch}{1.4}
\begin{table}[t]
\centering
\begin{tabular}{|c|c|c|c|}\hline\hline
$g_Z$                  & $T^2$ ${\rm (GeV^2)}$  & Values  & Corresponding decay widths {\rm (MeV)}              \\ \hline
$g_{\eta_cN,1}$        & $5.0-6.0$              & $2.89^{+0.04}_{-0.04}\,\rm{GeV^{-1}}$ &   $153.48^{+4.19}_{-4.19}$ \\ \hline
$g_{J/\psi N,2}$       & $6.5-7.5$              & $0.60^{+0.66}_{-0.60}$ &   $5.38^{+11.96}_{-5.38}$    \\ \hline
$g_{\eta_c\Delta,3}$   & $6.0-7.0$              & $2.01^{+0.70}_{-0.70}$ &   $55.63^{+39.07}_{-39.07}$    \\ \hline
$g_{J/\psi\Delta,4}$   & $6.5-7.5$              & $1.74^{+0.04}_{-0.04}$ &   $42.72^{+2.10}_{-2.10}$    \\ \hline
$g_{\eta_cN,5}$        & $5.0-6.0$              & $0.56^{+0.14}_{-0.14}\,\rm{GeV^{-1}}$ &   $6.24^{+3.05}_{-3.05}$    \\ \hline
$g_{J/\psi N,6}$       & $5.0-6.0$              & $0.83^{+0.08}_{-0.08}$ &   $11.13^{+2.06}_{-2.06}$    \\ \hline
$g_{\eta_c\Delta,7}$   & $6.0-7.0$              & $1.81^{+0.70}_{-0.70}$ &   $51.37^{+40.00}_{-40.00}$    \\ \hline
$g_{J/\psi\Delta,8}$   & $8.0-9.0$              & $2.01^{+1.56}_{-1.56}$ &   $75.51^{+117.12}_{-75.51}$    \\ \hline
$g_{\eta_cN,9}$        & $5.0-6.0$              & $1.26^{+0.06}_{-0.06}\,\rm{GeV^{-2}}$ &   $2.57^{+0.24}_{-0.24}$    \\ \hline
$g_{J/\psi N,10}$      & $6.0-7.0$              & $0.89^{+0.12}_{-0.12}\,\rm{GeV^{-2}}$ &   $3.06^{+0.82}_{-0.82}$    \\ \hline
$g_{\eta_c\Delta,11}$  & $6.0-7.0$              & $2.30^{+0.27}_{-0.27}\,\rm{GeV^{-1}}$ &   $17.32^{+3.99}_{-3.99}$    \\ \hline
$g_{J/\psi\Delta,12}$  & $5.0-6.0$              & $2.55^{+1.36}_{-1.31}\,\rm{GeV^{-1}}$ &   $60.22^{+63.98}_{-60.22}$    \\ \hline\hline
\end{tabular}
\caption{The hadronic coupling constants extracted from the Borel windows and their corresponding decay widths}\label{gconstants}
\end{table}

Until now, the strong decay constants could be numerically solved as $g_\mathcal{Y}=g_\mathcal{Y}(T_1^2,T_2^2)$ which rely on the corresponding free parameters $\mathcal{C}_{1\sim12}$, where $\mathcal{Y}$ represent the twelve different decay constants. One could determine the free parameters via the ``flat surfaces" of $g_\mathcal{Y}(T_1^2,T_2^2)$ among the Borel windows \cite{Myself-DcPc4312}. As is analyzed, it is straightforward to set $T_1^2=T_2^2=T^2$ for simplicity \cite{Decay-mole-WZG-WX,WZG-Z4600-decay,WZG-Pc4312-decay-tetra,DZGWang}. Furthermore, the Borel platforms of each coupling constant are determined under the same intervals of the Borel parameters $T^2$, set $\left(T^2\right)_{max}-\left(T^2\right)_{min}=1\,{\rm GeV^2}$. The free parameters are listed as,
\begin{eqnarray*}
\mathcal{C}_1&=&\left(-2.5140\times 10^{-4}+1.0559\times 10^{-5}T^2\right)\,\rm{GeV^{10}}\,,\nonumber \\
\mathcal{C}_2&=&\left(-1.6400\times 10^{-4}+5.0840\times 10^{-7}T^2\right)\,\rm{GeV^{9}}\,,\nonumber \\
\mathcal{C}_3&=&\left(-2.4000\times 10^{-4}-2.2080\times 10^{-6}T^2\right)\,\rm{GeV^{9}}\,,\nonumber \\
\mathcal{C}_4&=&\left(1.3250\times 10^{-6}+1.0998\times 10^{-7}T^2+3.5113\times10^{-8}T^4\right)\,\rm{GeV^{10}}\,,\nonumber \\
\mathcal{C}_5&=&\left(-3.6080\times 10^{-4}-2.4174\times 10^{-6}T^2\right)\,\rm{GeV^{10}}\,,\nonumber \\
\mathcal{C}_6&=&\left(-6.3488\times 10^{-6}+4.1775\times 10^{-7}T^2\right)\,\rm{GeV^{9}}\,,\nonumber \\
\mathcal{C}_7&=&\left(-3.6960\times 10^{-4}-3.3264\times 10^{-6}T^2+1.0349\times10^{-7}T^4\right)\,\rm{GeV^{9}}\,,\nonumber \\
\mathcal{C}_8&=&\left(-5.3440\times 10^{-4}+8.7642\times 10^{-6}T^2-4.1790\times10^{-7}T^4\right)\,\rm{GeV^{10}}\,,\nonumber \\
\mathcal{C}_9&=&\left(-5.0000\times 10^{-6}+2.5000\times 10^{-6}T^2+4.0000\times10^{-7}T^4\right)\,\rm{GeV^{11}}\,,\nonumber \\
\mathcal{C}_{10}&=&\left(2.0000\times 10^{-4}-3.4000\times 10^{-6}T^2\right)\,\rm{GeV^{10}}\,,\nonumber \\
\mathcal{C}_{11}&=&\left(-3.5405\times 10^{-6}+1.4516\times 10^{-7}T^2-3.8946\times10^{-8}T^4\right)\,\rm{GeV^{10}}\,,\nonumber \\
\mathcal{C}_{12}&=&\left(-1.2240\times 10^{-4}-7.3440\times 10^{-7}T^2+3.4272\times10^{-8}T^4\right)\,\rm{GeV^{9}}\,,\nonumber
\end{eqnarray*}
where, $T$ in the above expressions of $\mathcal{C}_{1\sim12}$ have no unit.
Just considering so many input parameters for the numerical calculation of the strong decay constants, their error bounds are somewhat complicated, the approximations $\frac{\delta \lambda_P}{\lambda_P}=\frac{\delta \lambda_{P^\prime}}{\lambda_{P^\prime}}=\frac{\delta \lambda_\Delta}{\lambda_\Delta}=\frac{\delta \lambda_N}{\lambda_N}=\frac{\delta f_{J/\psi}}{f_{J/\psi}}=\frac{\delta f_{\eta_c}}{f_{\eta_c}}$ are taken to estimate the uncertainties \cite{WZG-Pc4312-decay-tetra,DZGWang}, and the uncertainties of the $m_{P_{A\sim F}}$ are neglected to avoid over-evaluation. Furthermore, the error bounds due to the uncertainties of the free parameters $\delta \mathcal{C}_{1\sim 12}$ are abandoned \cite{Decay-mole-WZG-WX,WZG-Z4600-decay,WZG-Pc4312-decay-tetra,DZGWang}. Take the QCD sum rules of Eq.(49) for example, the uncertainty of the decay constant satisfies,
\begin{eqnarray}\label{Uncertainty}
&& \frac{f_{\eta_c}m_{\eta_c}^2\lambda_N\lambda_{P_A} g_{\eta_cN,1}}{2m_c \xi}\frac{\kappa_1}{\frac{m_{P_A}^2}{\xi}-m_{\eta_c}^2}\left\{ {\rm exp} \left( -\frac{m_{\eta_c}^2}{T_1^2} \right)-{\rm exp} \left( -\frac{m_{P_A}^2}{\xi T_1^2} \right) \right\}{\rm exp} \left( -\frac{m_N^2}{T_2^2} \right) \nonumber \\
&& \cdot \left(\frac{\delta f_{\eta_c}}{f_{\eta_c}}+\frac{\delta \lambda_N}{\lambda_N}+\frac{\delta\lambda_{P_A}}{\lambda_{P_A}}+\frac{\delta g_{\eta_cN,1}}{g_{\eta_cN,1}} \right)=\delta\int_{4m_c^2}^{s_{\eta_c}^0}ds\int_0^{s_N^0}du\, \rho_1(s,u){\rm exp}\left( -\frac{s}{T_1^2}-\frac{u}{T_2^2} \right)\,,
\end{eqnarray}
till now, there's no perfect way to deal with such complicated items like $\frac{\delta f_{\eta_c}}{f_{\eta_c}}+\frac{\delta \lambda_N}{\lambda_N}+\frac{\delta\lambda_{P_A}}{\lambda_{P_A}}+\frac{\delta g_{\eta_cN,1}}{g_{\eta_cN,1}}$ to find a practical way to solve to uncertainty of the strong decay constant strictly, the approximation $\frac{\delta f_{\eta_c}}{f_{\eta_c}}=\frac{\delta \lambda_N}{\lambda_N}=\frac{\delta\lambda_{P_A}}{\lambda_{P_A}}=\frac{\delta g_{\eta_cN,1}}{g_{\eta_cN,1}}$ is applied, the uncertainty of $g_{\eta_cN,1}$ is derived as,
\begin{eqnarray}
\delta g_{\eta_cN,1}&=&\frac{1}{4}\sqrt{\sum_i\left[g_{\eta_cN,1}(x_i+\delta x_i)-g_{\eta_cN,1}(x_i)\right]^2},\nonumber
\end{eqnarray}
where, $x_i$ are the input parameters due to the vacuum condensates $\langle\overline{q}q\rangle$, $\langle\overline{q}g_s\sigma Gq\rangle$ and $m_c(m_c)$, $\delta x_i$ are their uncertainties. The reason to neglect the uncertainties of masses to avoid over-evaluation, such like $m_{P_A\sim F}$, $m_{\eta_c}$ and so on, is then obvious, since their uncertainties are also from these input parameters $x_i$ under the framework of QCD sum rules.

Accordingly, the graphs of the decay constants are shown in the Figures $1\sim6$, and the extracted values of the strong decay constants are listed in the Table 1, and the related decay widths are,

\begin{eqnarray}
\Gamma_{J,1\sim12}&=&\sum_s |\mathcal{T}_{1\sim12}^2|\frac{\mathcal{S}(m_{P_{A\sim F}},m_1,m_2)}{8(2J+1)\pi m_{P_{A\sim F}}^2}
\end{eqnarray}
where, $J$ is the angular momentum of the states $\mathcal{P}_{A\sim F}$, $m_1$ and $m_2$ are the masses of the decay products, $\mathcal{S}(a,b,c)=\frac{\sqrt{a^2-(b+c)^2}\sqrt{a^2-(b-c)^2}}{2a}$, $\mathcal{T}_{1\sim12}$ are the decay vertices listed on the left sides of Eqs. [41-46]. Take the decay channel $P_c(4380)\rightarrow \eta_cN$ for example, $m_1=m_{\eta_c}$, $m_2=m_{N}$, $J=\frac{3}{2}$ and $\mathcal{T}_1=\langle \eta_c(p)N(q)| \mathcal{P}_A(p')\rangle$. The decay widths $\Gamma_{A\sim F}$ of the $\mathcal{P}_{A\sim F}$ are determined as,
\begin{eqnarray*}
\Gamma_A&=&158.86^{+12.67}_{\,\,-6.82}\,\rm{MeV}\,,\\
\Gamma_B&=&98.35^{+39.13}_{-39.13}\,\rm{MeV}\,,\\
\Gamma_C&=&17.37^{+3.68}_{-3.68}\,\rm{MeV}\,,\\
\Gamma_D&=&126.88^{+123.76}_{\,\,-85.45}\,\rm{MeV}\,,\\
\Gamma_E&=&5.63^{+0.85}_{-0.85}\,\rm{MeV}\,,\\
\Gamma_F&=&77.54^{+64.10}_{-60.35}\,\rm{MeV}\,.
\end{eqnarray*}
The ratios of the partial decay widths of the $\mathcal{P}_{A\sim F}$ are
\begin{eqnarray*}
\frac{\Gamma[P_c(4380)\rightarrow \eta_cN]}{\Gamma[P_c(4380)\rightarrow J/\psi N]}&=&28.53\,,\\
\frac{\Gamma[P_c(4410)\rightarrow \eta_c\Delta]}{\Gamma[P_c(4410)\rightarrow J/\psi \Delta]}&=&1.30\,,\\
\frac{\Gamma[P_c(4440)\rightarrow \eta_cN]}{\Gamma[P_c(4440)\rightarrow J/\psi N]}&=&0.56\,,\\
\frac{\Gamma[P_c(4470)\rightarrow \eta_c\Delta]}{\Gamma[P_c(4470)\rightarrow J/\psi \Delta]}&=&0.68\,,\\
\frac{\Gamma[P_c(4457)\rightarrow \eta_cN]}{\Gamma[P_c(4457)\rightarrow J/\psi N]}&=&0.84\,,\\
\frac{\Gamma[P_c(4620)\rightarrow \eta_c\Delta]}{\Gamma[P_c(4620)\rightarrow J/\psi \Delta]}&=&0.29\,.
\end{eqnarray*}

Results show that the decay widths of the discovered $P_c(4380)$, $P_c(4440)$ and $P_c(4457)$ are in good agreement with data proposed by the LHCb \cite{RAaij1,RAaij2}. Especially, for the wide width `$P_c(4380)$' state, the main decay channel is $P_c(4380)\rightarrow \eta_cN$, it contributes nearly $97\%$ among the wide width $158.86\,\rm{MeV}$ which fits the data $205\pm18\pm86\,\rm{MeV}$ observed by the LHCb.

The value of $\xi$ affects the value of the correlation functions at both the hadronic side and QCD side, and therefor, the chosen the related free parameter $\mathcal{C}$ and Borel platform. As the approximation model of our previous work, we directly set $p'^2=p^2$. Based on the range of $\xi$, $0\leq \xi\leq \frac{2m_N^2}{m_{\eta_c}^2}+2$ and considering $\frac{m_N^2}{m_{\eta_c^2}}\approx 0.1$, say, for the decay channel $P_c(4380)\rightarrow\eta_c+N$, it is more reasonable to set $\xi=\frac{m_N^2}{m_{\eta_c^2}}+1$. To argue the stability of the effect of $\xi$, we make a comparison for these two approximation models. As shown in the Fig. 7, the value of the hadronic coupling constant extracted from the Borel window is $g_{\eta_cN,1}^{\prime}=2.91^{+0.04}_{-0.04}\,\,\rm{GeV^{-1}}$ with $\xi^\prime=1$ and the free parameter $\mathcal{C}_1^\prime=-2.7490\times 10^{-4}+1.0721\times 10^{-5}T^2\,\,\rm{GeV^{10}}$, results show that, slight change occurs for the hadronic coupling constant, but the stability is robust.

In the consideration of the two Borel parameters $T_1^2$ and $T_2^2$, the chosen Borel platform should be a `flat' plane \cite{Myself-DcPc4312}. Thus, any `straight lines' on the plane are also flat, it is reasonable to simply set $T_1^2=T_2^2$ without significant change of the Borel window and the resulting value of the hadronic coupling constant. Of course, this may cause a slight difference. In order to make a clear discussion about this, we still take the decay channel $P_c(4380)\rightarrow\eta_c+N$ for example to make the comparison. In the Fig. 8, the 3-D graph of the hadronic coupling constant $g^{\prime\prime}_{\eta_cN,1}$ of the decay channel $P_c(4380)\rightarrow \eta_cN$ among the Borel window is shown, the extracted value of the hadronic coupling constant is $g^{\prime\prime}_{\eta_cN,1}=2.87^{+0.05}_{-0.04}\,\,\rm{GeV^{-1}}$ with the free parameter $\mathcal{C}_1^{\prime\prime}=-2.5000\times 10^{-4}+1.0250\times 10^{-5}T_1^2+2.5000\times 10^{-7}T_2^2\,\,\rm{GeV^{10}}$, the difference is acceptable.

Need to point out that the chosen vacuum condensates are $\langle\overline{q}q\rangle$, $\langle\frac{\alpha_s}{\pi}GG\rangle$, $\langle\overline{q}g_s\sigma Gq\rangle$, $\langle\overline{q}q\rangle^2$, $g_s^2\langle\overline{q}q\rangle^2$, $\langle\frac{\alpha_s}{\pi}GG\rangle\langle\overline{q}q\rangle$, $\langle\overline{q}q\rangle\langle\overline{q}g_s\sigma Gq\rangle$, $g_s^2\langle\overline{q}q\rangle^3$, $\langle\frac{\alpha_s}{\pi}GG\rangle\langle\overline{q}q\rangle^2$ and $\langle\overline{q}g_s\sigma Gq\rangle^2$ on the QCD sides. In the framework of QCD sum rules, these vacuum condensates are parameterized, and their values rely on the energy scale $\mu$, thus, the results in the present work have the energy dependence. The decay products of all the decay channels studied now are the conventional hadrons. For the coupling of the QCD side and hadron side of the conventional hadron channel, there is no unified standard to determined the energy scale, it is reasonable to determine the energy at an acceptable range. For example, we set the energy scale $\mu=1\,\rm{GeV}$ to study the strong decay of hidden-charm tetraquark state candidate $X(3872)$ and $Y(4500)$ \cite{WZG-X3872,WZG-Y4500}, while in Refs. \cite{Myself-DcPc4312,Myself-DcPcs4338} the energy scale is set as $\frac{m_{\eta_c}}{2}\,\rm{GeV}$ or $\frac{m_{J/\psi}}{2}\,\rm{GeV}$ which is acceptable for the  charmonium states \cite{EWZGHuang}, and in Ref. \cite {WZG-X4140-decay}, the energy scale is chosen as $\mu=2\,\rm{GeV}$ to calculate the strong decay of $X(4140)$ \cite{WZG-X4140-decay}. We plot the energy scale dependence of the vacuum condensates $\langle\overline{q}q\rangle$ and $\langle\overline{q}g_s\sigma Gq\rangle$ in the Fig. [9], one can find that the value of these two vacuum condensates do not have too much difference with the energy scales varying from $1\,\rm{GeV}$ to $2\,\rm{GeV}$, their tiny difference are less than their corresponding error bounds. The difference mainly occur for the strong fine-structure constant $\alpha_s$ and $m_c$, $g_s^2=6.7381$ and $3.6850$ at energy scale $\mu=1\,\rm{GeV}$ and $2\,\rm{GeV}$, respectively. The four-quark condensate $g_s^2\langle \bar{q}q\rangle^2$ comes from the terms
$\langle \bar{q}\gamma_\mu  q g_s D_\eta G_{\lambda\tau}\rangle$, $\langle\bar{q}_jD^{\dagger}_{\mu}D^{\dagger}_{\nu}D^{\dagger}_{\alpha}q_i\rangle$  and
$\langle\bar{q}_jD_{\mu}D_{\nu}D_{\alpha}q_i\rangle$  rather than comes from the radiative  $\mathcal{O}(\alpha_s)$ corrections for the four-quark condensate $\langle \bar{q}q\rangle^2$, where $D_\alpha=\partial_\alpha-ig_sG_\alpha $. The strong coupling constant $\alpha_s(\mu)=\frac{g_s^2(\mu)}{4\pi}$ appears  at the tree level. In fact, the contributions of such terms are tiny. Thus, there is not too much energy scale dependence for the contribution of the vacuum condensates.
Taking the decay channel $P_c(4380)\rightarrow\eta_c+N$ for example, the energy scale dependence are mainly from following lower limit of integral
\begin{eqnarray*}
&&\int_{4m_c^2}^{s_{\eta_c}^0}ds\int_0^{s_N^0}du\, \rho_1(s,u){\rm exp}\left( -\frac{s}{T_1^2}-\frac{u}{T_2^2} \right)\,.
\end{eqnarray*}
Since $4m_c^2=8.8659\,\rm{GeV^2}$ and $4.9671\,\rm{GeV^2}$ at energy scale $\mu=1\,\rm{GeV}$ and $2\,\rm{GeV}$, respectively. In the present work, we follow our previous work to set a medium energy scale $\mu=\frac{m_{\eta_c}}{2}\,\rm{GeV}$ or $\mu=\frac{m_{J/\psi}}{2}\,\rm{GeV}$ \cite{Myself-DcPc4312,Myself-DcPcs4338} to avoid relatively large energy scale for the conventional hadron, and to ensure the integral range to maintain the stability of the Borel windows.

Before the present work, several theoretic groups had proposed different physical models to study the strong decays of the discovered $P_c$ states \cite{Decay-mole-Sakai,Decay-mole-Zou,Decay-mole-HeChen,Decay-mole-GJWang,Decay-mole-Gutsche,Decay-mole-WZG-WX,Decay-mole-HuangMQ,Pcdecay0,Pcdecay01,Pcdecay02,Pcdecay1,Pcdecay2,Pcdecay3}. For example, in Ref. \cite{Pcdecay1}, the $P_c(4440)$ and $P_c(4457)$ are assigned as the $\Sigma_c\bar{D}^*$ hadronic molecule with $J^P$ being $\frac{1}{2}^-$ or $\frac{3}{2}^-$ via the heavy quark spin symmetry (HQSS), for $\frac{1}{2}^-$, $\frac{\Gamma[P_c\rightarrow\eta_cp]}{\Gamma[P_c\rightarrow J/\psi p]}=\frac{3}{25}$, furthermore, $\frac{\Gamma[P_c\rightarrow\Lambda_c\bar{D}^*]}{\Gamma[P_c\rightarrow\Lambda_c\bar{D}]}=\frac{4}{3}$. By combining the effective Lagrangian and Bethe-Salpeter framework, Ref. \cite{Pcdecay2} studied the strong decays of $P_c(4440)$ and $P_c(4457)$ for the decay channels $\bar{D}^{(*)0}\Lambda_c^+$, $J/\psi(\eta_c)p$ and $\bar{D}\Sigma_c^{(*)}$, it is suggested that the $\bar{D}^{0}\Lambda_c$ and $\bar{D}^{*0}\Lambda_c$ are the dominant decay channels for these two $P_c$ states. Considering the compact pentaquark picture of the $uudc\bar{c}$ pentaquark based on the flavor-spin model extended to $SU(4)$, Ref. \cite{Pcdecay3} studied the ratios of decay rates of the $P_c$ pentaquarks to $J/\psi$ and similarly the ratios of decay rates to $\Lambda_c\bar{D}^*$ and $\Lambda_c\bar{D}$, it is proposed that $\frac{\Gamma[P_c(\frac{1}{2}^-)\rightarrow\eta_cp]}{\Gamma[P_c(\frac{1}{2}^-)\rightarrow J/\psi p]}=\frac{1}{3}$. The fact is that arguments for the strong decays of these $P_c$ states have divergences proposed by different theoretic groups. It is hard to claim which results having the discriminating power among the other different scenarios, of course, including the predictions made in the present study. We noticed the very resent work of LHCb \cite{Pcdecay4}, the collaboration performed a search for $P_c(4312)$, $P_c(4440)$ and $P_c(4457)$ states in the prompt $\Sigma_c\bar{D}^{(*)}$, $\Sigma_c^*\bar{D}^{(*)}$, $\Lambda_c^+\bar{D}^{(*)}$ and $\Lambda_c^+ \pi \bar{D}^{(*)}$ mass spectra, and their signal yields are found to be consistent with zero in all cases, thus, we follow this experimental results and study the present decay channels.

To date, we have systematically studied the hidden-charm pentaquark state candidates $P_c(4312)$, $P_c(4380)$, $P_c(4440)$ and $P_c(4457)$, in Ref. \cite{mass-mole-WXW-SCPMA}, the isospin is clearly differentiated for the first time under the framework of QCD sum rules, they are assigned as the $\bar{D}\Sigma_c$, $\bar{D}\Sigma_c^*$, $\bar{D}^*\Sigma_c$ and $\bar{D}^*\Sigma_c^*$ molecular states with low isospin $I=\frac{1}{2}$, their $J^P$  are $\frac{1}{2}^-$, $\frac{3}{2}^-$, $\frac{3}{2}^-$ and $\frac{5}{2}^-$, respectively. Under the heated debates for the nature of these $P_c$ states, we propose our own reasonable arguments. More significantly, the high isospin $I=\frac{3}{2}$ cousins $P_c(4330)$, $P_c(4410)$, $P_c(4470)$ and $P_c(4620)$ are predicted. Results of the present work show that these high isospin cousins have relatively wide widths which are reasonable for assigning them as resonance states. The predictions for the high isospin cousins would provide an explicit reference for the future experiments and will testify our interpretation for the nature of the observed $P_c(4312)$, $P_c(4380)$, $P_c(4440)$ and $P_c(4457)$ in return.

\section{Conclusions}
In the present work, the strong decays of the exotic $P_c(4380)$, $P_c(4440)$, $P_c(4457)$ and their possible isospin cousins are studied. The decay channels of these $P_c$ states $\mathcal{P}_{A\sim F}\rightarrow \eta_c(J/\psi)+N(\Delta)$ are calculated via the QCD sum rules. The decay widths of the $P_c(4380)$, $P_c(4440)$ and $P_c(4457)$ derived in this study are in good agreement with the experimental results observed by the LHCb. The predicted strong decay widths for the $P_c(4410)$, $P_c(4470)$ and $P_c(4620)$ are waiting for the verification of the future experiments. Moreover, the ratios of the partial decay widths of these $P_c$ states are solved, for example, it is found that the main decay channel for the $P_c(4380)$ is $P_c(4380)\rightarrow\eta_cN$, these theoretical results may provide a reference for the high energy experiments. The strong decays of $P_{cs}(4338)$ had been analyzed by our group \cite{Myself-DcPcs4338}, the strong decays of $P_{cs}(4459)$ and $P_c(4337)$ would be our next target in the near future for the systematic picture of the strong decays of the pentaquark exotic $P_c$ and $P_{cs}$ states.

\section*{Acknowledgements}
This work is supported by the National Natural Science Foundation with Grant Number 12575083 and the Natural Science Foundation of Hebei Province with Grant Number A2024502002.

\section*{Appendix}
\begin{eqnarray}
\kappa_2&=&\frac{4(m_{J/\psi}^2 \tau+m_N^2)}{3 m_{J/\psi}^2 \xi}+\frac{8 m_N m_{P_A}}{3 m_{J/\psi}^2 \xi}
\end{eqnarray}

\begin{eqnarray}
\kappa_3&=&-\frac{4 m_{\eta_c}^2 \tau^2}{9 m_{\Delta}^2 \xi}+\frac{4 m_{\Delta}^2}{9 m_{\eta_c}^2 \xi}+\frac{8 m_{\eta_c}^2 \tau}{9 m_{\Delta}^2}+\frac{4 m_{\Delta} m_{P_B}}{3 m_{\eta_c}^2 \xi}\nonumber\\
&&-\frac{4 m_{\eta_c}^2 \xi}{9 m_{\Delta} m_{P_B}}+\frac{4 m_{\eta_c}^2 \tau}{9 m_{\Delta} m_{P_B}}-\frac{4 m_{P_B} \tau}{3 m_{\Delta} \xi}+\frac{8 m_{P_B}}{3 m_{\Delta}}+\frac{16}{9}
\end{eqnarray}

\begin{eqnarray}
\kappa_4&=&-\frac{8 m_{\Delta}^2 m_{P_B}}{9 m_{J/\psi}^2 \xi}+\frac{2 m_{\Delta}^2}{9 m_{P_B}}+\frac{2 m_{\Delta}^3}{9 m_{J/\psi}^2 \xi}+\frac{2 m_{J/\psi}^2 \tau^2}{9 m_{\Delta} \xi}-\frac{4 m_{J/\psi}^2 \tau}{9 m_{\Delta}}\nonumber\\
&&+\frac{4 m_{\Delta} \tau}{9 \xi}-\frac{4 m_{\Delta}}{9}-\frac{4 m_{J/\psi}^2 \xi}{9 m_{P_B}}+\frac{2 m_{J/\psi}^2 \tau}{9 m_{P_B}}-\frac{8 m_{P_B} \tau}{9 \xi}+\frac{16 m_{P_B}}{9}
\end{eqnarray}

\begin{eqnarray}
\kappa_5&=&\frac{4 m_N^3 m_{P_C}}{3 m_{\eta_c}^2 \xi}-\frac{2 m_N^4}{3 m_{\eta_c}^2 \xi}+\frac{4 m_N^2}{3 \xi}-\frac{4 m_N m_{P_C} \tau}{3 \xi}-\frac{8 m_N m_{P_C}}{3 \xi}+\frac{2 m_{\eta_c}^2 \tau^2}{3 \xi}+\frac{4 m_{\eta_c}^2 \tau}{3 \xi}
\end{eqnarray}

\begin{eqnarray}
\kappa_6&=&\frac{4 m_N^2}{3 m_{J/\psi}^2 \xi}+\frac{8 m_N m_{P_C}}{3 m_{J/\psi}^2 \xi}+\frac{4 \tau}{3 \xi}
\end{eqnarray}

\begin{eqnarray}
\kappa_7&=&\frac{4 m_{\eta_c}^2 \tau^2}{9 m_{\Delta}^2 \xi}-\frac{4 m_{\Delta}^2}{9 m_{\eta_c}^2 \xi}-\frac{8 m_{\eta_c}^2 \tau}{9 m_{\Delta}^2}-\frac{4 m_{\Delta} m_{P_D}}{3 m_{\eta_c}^2 \xi}\nonumber\\
&&+\frac{4 m_{\eta_c}^2 \xi}{9 m_{\Delta} m_{P_D}}-\frac{4 m_{\eta_c}^2 \tau}{9 m_{\Delta} m_{P_D}}+\frac{4 m_{P_D} \tau}{3 m_{\Delta} \xi}-\frac{8 m_{P_D}}{3 m_{\Delta}}-\frac{16}{9}
\end{eqnarray}

\begin{eqnarray}
\kappa_8&=&\frac{8 m_{\Delta}^2 m_{P_D}}{9 m_{J/\psi}^2 \xi}-\frac{2 m_{\Delta}^2}{9 m_{P_D}}-\frac{2 m_{\Delta}^3}{9 m_{J/\psi}^2 \xi}-\frac{2 m_{J/\psi}^2 \tau^2}{9 m_{\Delta} \xi}+\frac{4 m_{J/\psi}^2 \tau}{9 m_{\Delta}}\nonumber\\
&&-\frac{4 m_{\Delta} \tau}{9 \xi}+\frac{4 m_{\Delta}}{9}+\frac{4 m_{J/\psi}^2 \xi}{9 m_{P_D}}-\frac{2 m_{J/\psi}^2 \tau}{9 m_{P_D}}+\frac{8 m_{P_D} \tau}{9 \xi}-\frac{16 m_{P_D}}{9}
\end{eqnarray}

\begin{eqnarray}
\kappa_9&=&\frac{m_N^5 m_{P_E}}{5 m_{\eta_c}^2 \xi}-\frac{2 m_N^3 m_{P_E} \tau}{5 \xi}-\frac{4 m_N^3 m_{P_E}}{5 \xi}-\frac{m_N^2 m_{\eta_c}^2 \tau^2}{10 \xi}+\frac{m_N^6}{10 m_{\eta_c}^2 \xi}\nonumber\\
&&+\frac{2 m_N^2 m_{\eta_c}^2}{5 \xi}-\frac{2 m_N^2 m_{\eta_c}^2}{5}-\frac{m_N^4 \tau}{10 \xi}-\frac{2 m_N^4}{5 \xi}+\frac{m_N m_{P_E} m_{\eta_c}^2 \tau^2}{5 \xi}+\frac{4 m_N m_{P_E} m_{\eta_c}^2 \tau}{5 \xi}\nonumber\\
&&+\frac{4 m_N m_{P_E} m_{\eta_c}^2}{5 \xi}-\frac{4}{5} m_N m_{P_E} m_{\eta_c}^2+\frac{m_{\eta_c}^4 \tau^3}{10 \xi}+\frac{2 m_{\eta_c}^4 \tau^2}{5 \xi}+\frac{2 m_{\eta_c}^4 \tau}{5 \xi}-\frac{2 m_{\eta_c}^4 \tau}{5}
\end{eqnarray}

\begin{eqnarray}
\kappa_{10}&=&\frac{m_N^4 m_{P_E}}{5 m_{J/\psi}^2 \xi}-\frac{2 m_N^2 m_{P_E} \tau}{5 \xi}-\frac{4 m_N^2 m_{P_E}}{5 \xi}-\frac{m_N^5}{5 m_{J/\psi}^2 \xi}+\frac{2 m_N^3 \tau}{5 \xi}+\frac{4 m_N^3}{5 \xi}\nonumber\\
&&-\frac{m_N m_{J/\psi}^2 \tau^2}{5 \xi}-\frac{4 m_N m_{J/\psi}^2 \tau}{5 \xi}-\frac{4 m_N m_{J/\psi}^2}{5 \xi}+\frac{4 m_N m_{J/\psi}^2}{5}+\frac{m_{P_E} m_{J/\psi}^2 \tau^2}{5 \xi}\nonumber\\
&&+\frac{4 m_{P_E} m_{J/\psi}^2 \tau}{5 \xi}+\frac{4 m_{P_E} m_{J/\psi}^2}{5 \xi}-\frac{4 m_{P_E} m_{J/\psi}^2}{5}
\end{eqnarray}

\begin{eqnarray}
\kappa_{11}&=&\frac{2 m_{\Delta}^3 m_{P_F}}{15 m_{\eta_c}^2 \xi}+\frac{2 m_{\Delta}^2 m_{P_F}}{15 m_{\eta_c} \sqrt{\xi}}-\frac{m_{\eta_c}^4 \tau^3}{15 m_{\Delta}^2 \xi}-\frac{2 m_{\eta_c}^4 \tau^2}{15 m_{\Delta}^2 \xi}-\frac{2 m_{\Delta}^4}{15 m_{\eta_c}^2 \xi}\nonumber\\
&&+\frac{2 m_{\eta_c}^4 \tau^2}{15 m_{\Delta}^2}+\frac{2 m_{\Delta}^3}{15 m_{\eta_c} \sqrt{\xi}}+\frac{m_{\Delta}^2 \tau}{15 \xi}+\frac{4 m_{\Delta}^2}{15 \xi}-\frac{4 m_{\Delta}^2}{15}+\frac{2 m_{P_F} m_{\eta_c}^2 \tau^2}{15 m_{\Delta} \xi}\nonumber\\
&&+\frac{4 m_{P_F} m_{\eta_c}^2 \tau}{15 m_{\Delta} \xi}-\frac{4 m_{P_F} m_{\eta_c}^2 \tau}{15 m_{\Delta}}-\frac{4 m_{\Delta} m_{P_F} \tau}{15 \xi}-\frac{4 m_{\Delta} m_{P_F}}{15 \xi}+\frac{4 m_{\Delta} m_{P_F}}{15}\nonumber\\
&&-\frac{2 m_{\Delta} m_{\eta_c} \tau}{15 \sqrt{\xi}}-\frac{4 m_{\Delta} m_{\eta_c}}{15 \sqrt{\xi}}-\frac{2 m_{P_F} m_{\eta_c} \tau}{15 \sqrt{\xi}}-\frac{4 m_{P_F} m_{\eta_c}}{15 \sqrt{\xi}}+\frac{2 m_{\eta_c}^2 \tau^2}{15 \xi}\nonumber\\
&&+\frac{2 m_{\eta_c}^2 \tau}{15 \xi}+\frac{2 m_{\eta_c}^2 \tau}{15}+\frac{4 m_{\eta_c}^2}{15}
\end{eqnarray}

\begin{eqnarray}
\kappa_{12}&=&-\frac{2 m_{P_F} m_{J/\psi}^2 \tau^2}{15 m_{\Delta}^2 \xi}-\frac{4 m_{P_F} m_{J/\psi}^2 \tau}{15 m_{\Delta}^2 \xi}+\frac{4 m_{P_F} m_{J/\psi}^2 \tau}{15 m_{\Delta}^2}-\frac{2 m_{J/\psi}^2 \tau^2}{15 m_{\Delta} \xi}\nonumber\\
&&-\frac{4 m_{J/\psi}^2 \tau}{15 m_{\Delta} \xi}+\frac{4 m_{J/\psi}^2 \tau}{15 m_{\Delta}}+\frac{2 m_{\Delta} \tau}{15 \xi}-\frac{4 m_{\Delta}}{15}+\frac{2 m_{P_F} \tau}{15 \xi}-\frac{4 m_{P_F}}{15}
\end{eqnarray}
\clearpage
\begin{figure}
 \centering
 \includegraphics[totalheight=5cm,width=7cm]{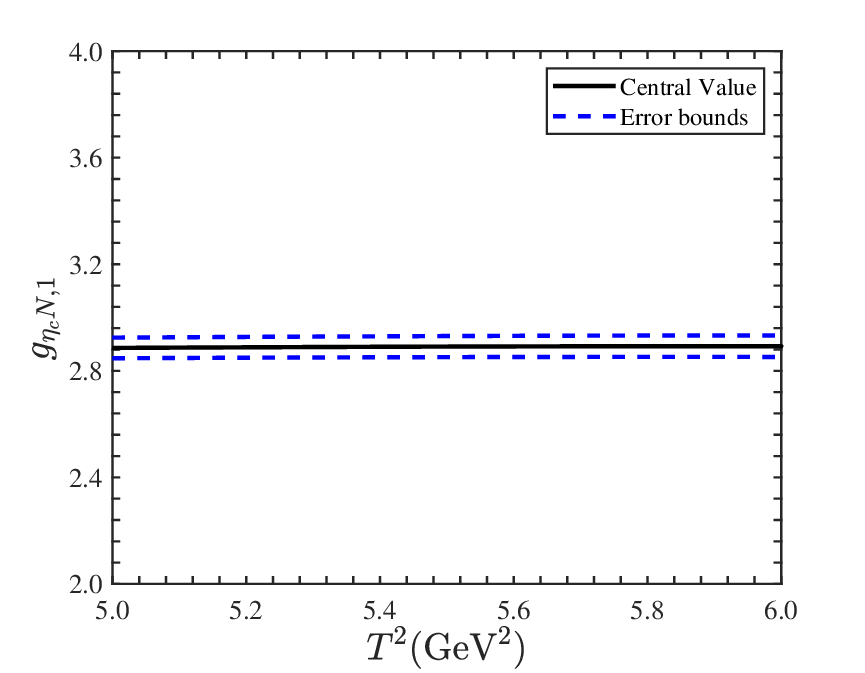}
 \includegraphics[totalheight=5cm,width=7cm]{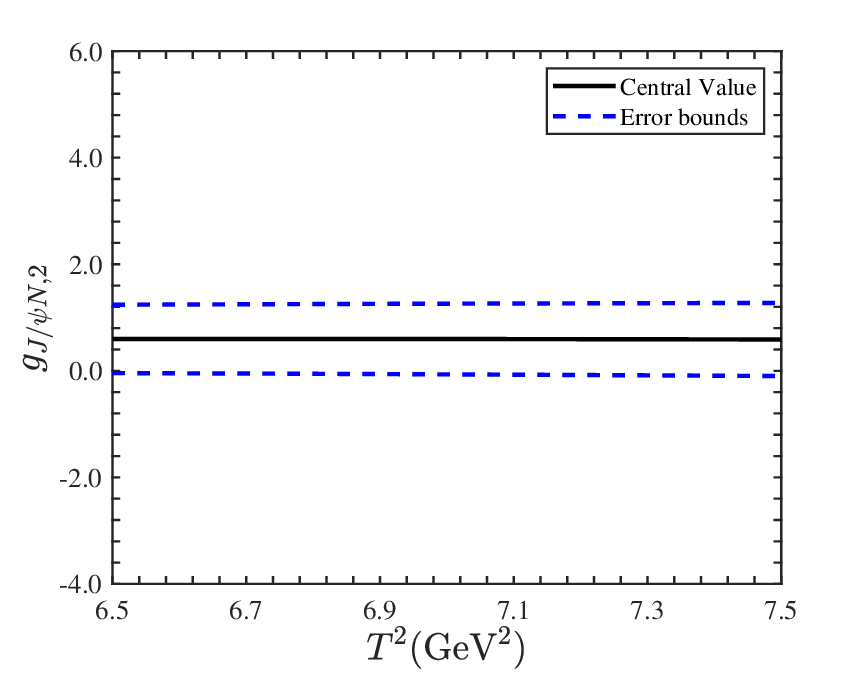}
 \caption{The hadronic coupling constants for $P_c(4380)\rightarrow \eta_cN$ (Left) and $P_c(4380)\rightarrow J/\psi N$ (Right) among the Borel windows.}\label{baryon-fig}
\end{figure}
\begin{figure}
 \centering
 \includegraphics[totalheight=5cm,width=7cm]{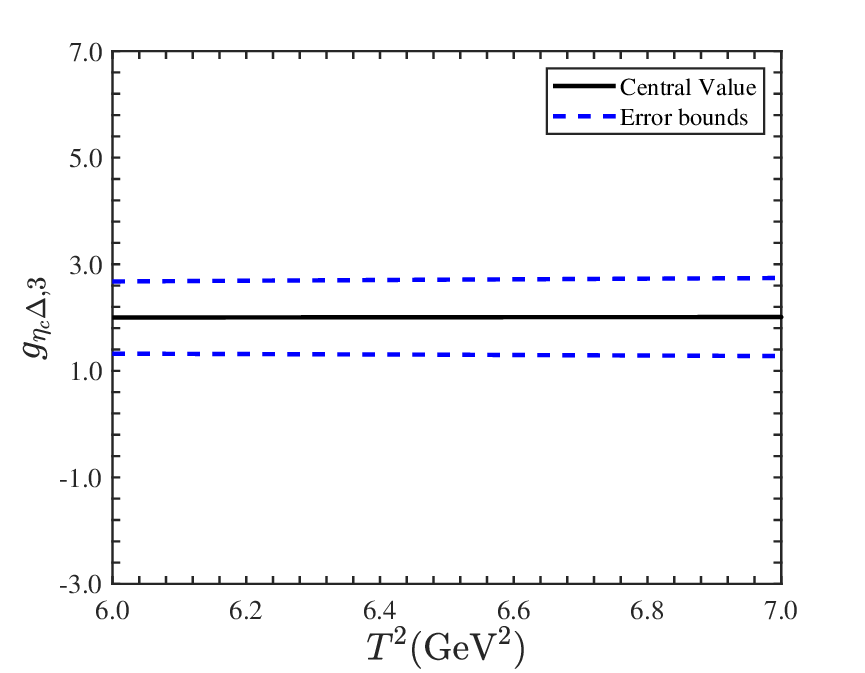}
 \includegraphics[totalheight=5cm,width=7cm]{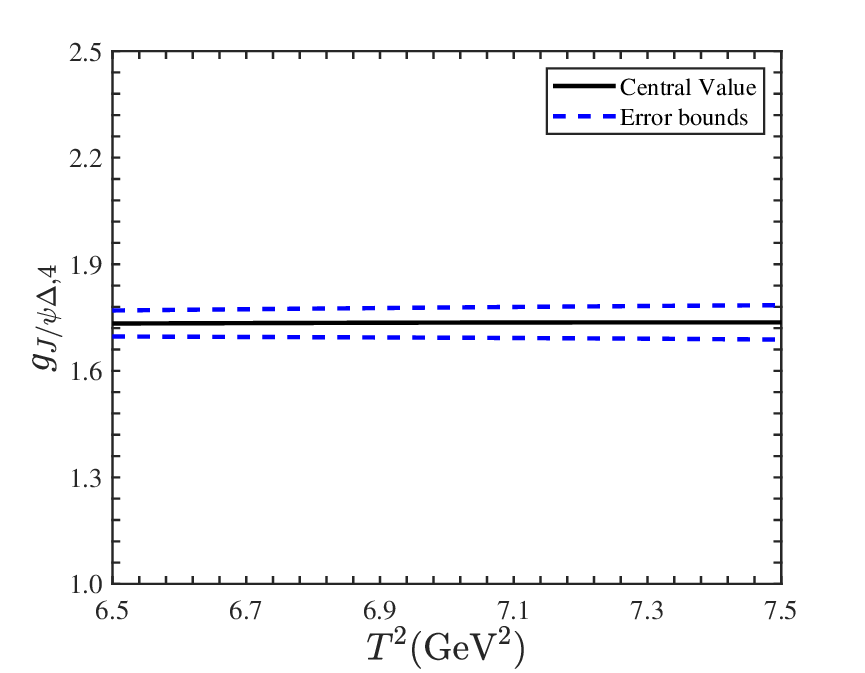}
 \caption{The hadronic coupling constants for $P_c(4410)\rightarrow \eta_c\Delta$ (Left) and $P_c(4410)\rightarrow J/\psi\Delta$ (Right) among the Borel windows.}\label{baryon-fig}
\end{figure}
\begin{figure}
 \centering
 \includegraphics[totalheight=5cm,width=7cm]{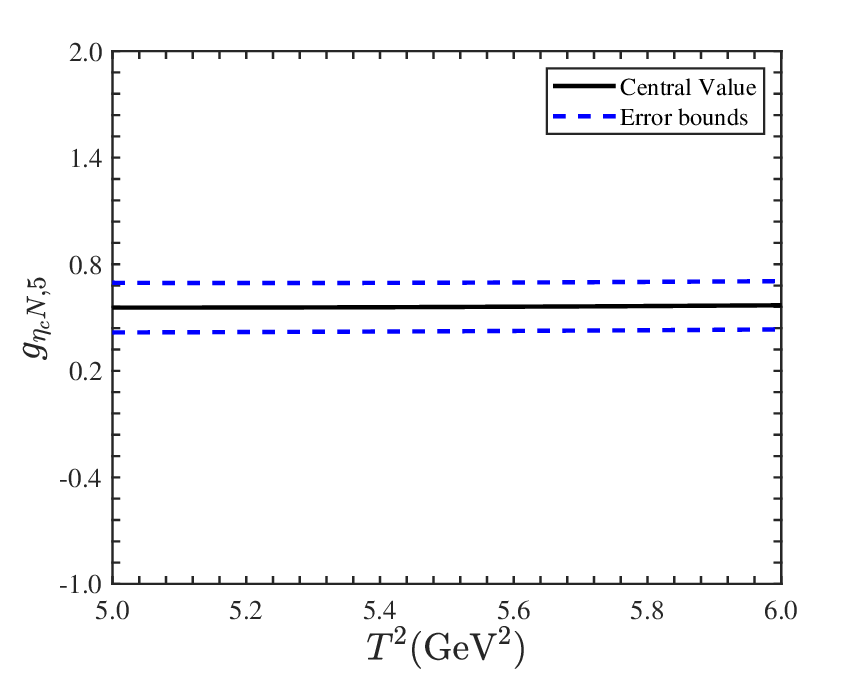}
 \includegraphics[totalheight=5cm,width=7cm]{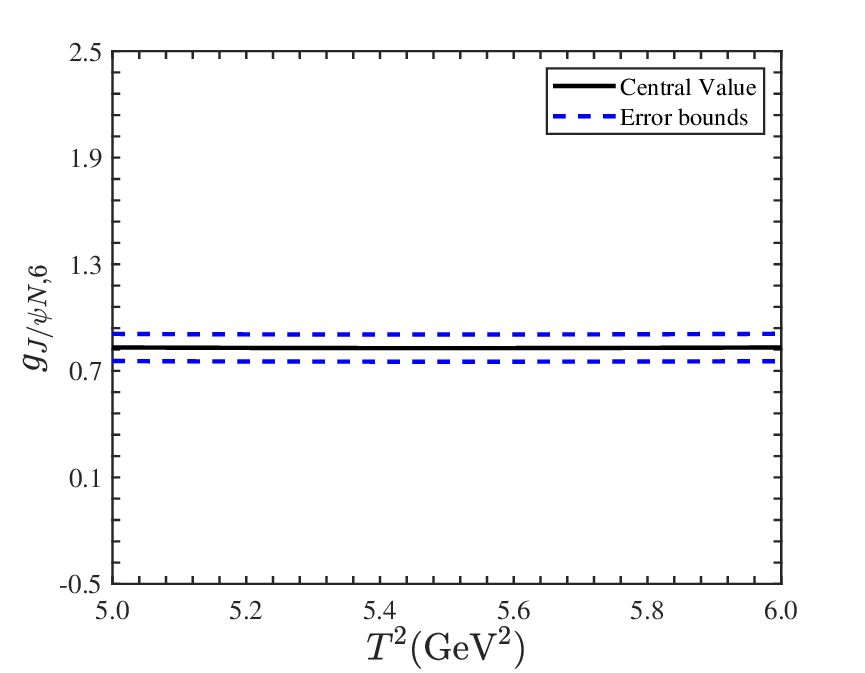}
 \caption{The hadronic coupling constants for $P_c(4440)\rightarrow \eta_cN$ (Left) and $P_c(4440)\rightarrow J/\psi N$ (Right) among the Borel windows.}\label{baryon-fig}
\end{figure}

\begin{figure}
 \centering
 \includegraphics[totalheight=5cm,width=7cm]{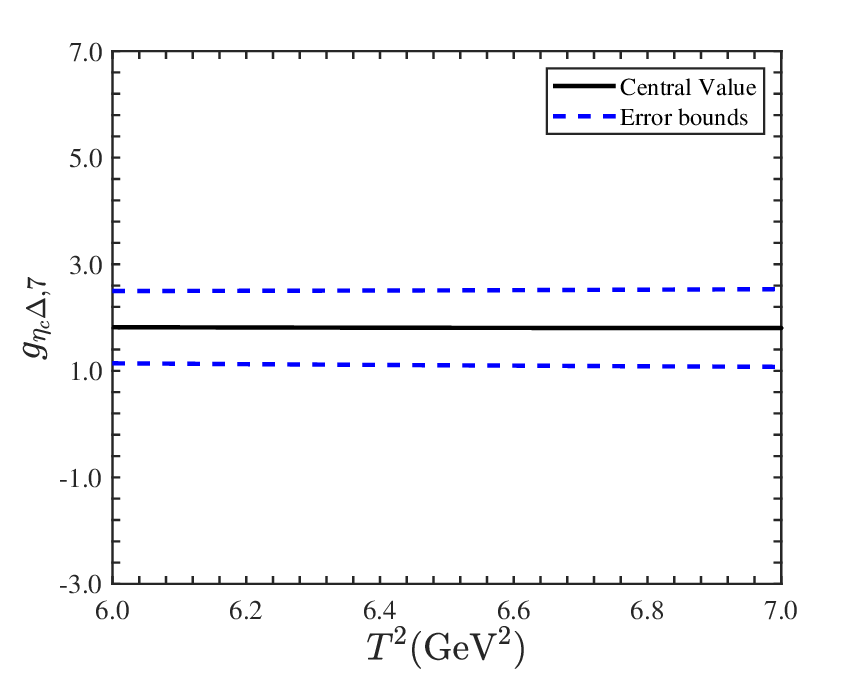}
 \includegraphics[totalheight=5cm,width=7cm]{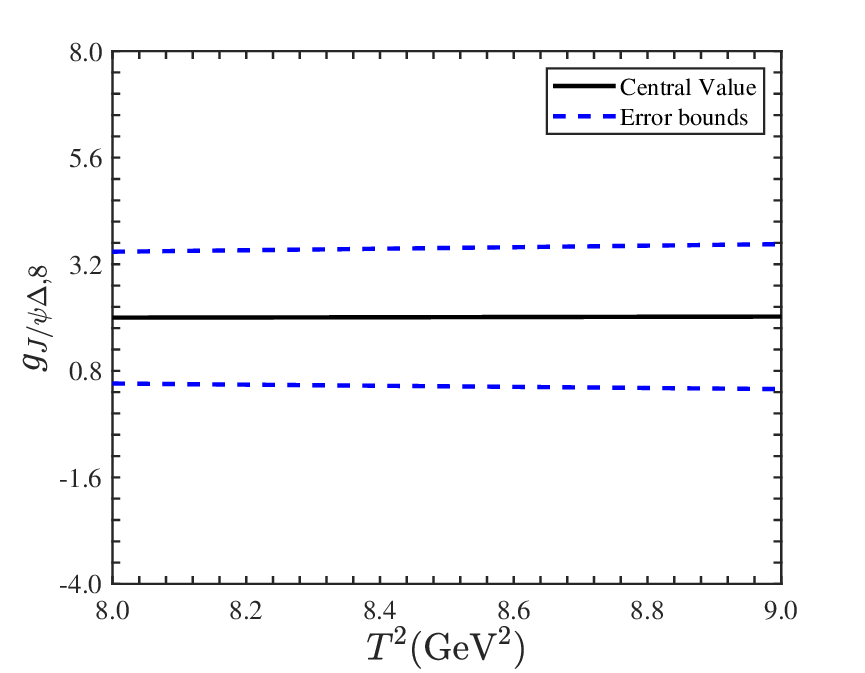}
 \caption{The hadronic coupling constants for $P_c(4470)\rightarrow \eta_c\Delta$ (Left) and $P_c(4470)\rightarrow J/\psi\Delta$ (Right) among the Borel windows.}\label{baryon-fig}
\end{figure}
\begin{figure}
 \centering
 \includegraphics[totalheight=5cm,width=7cm]{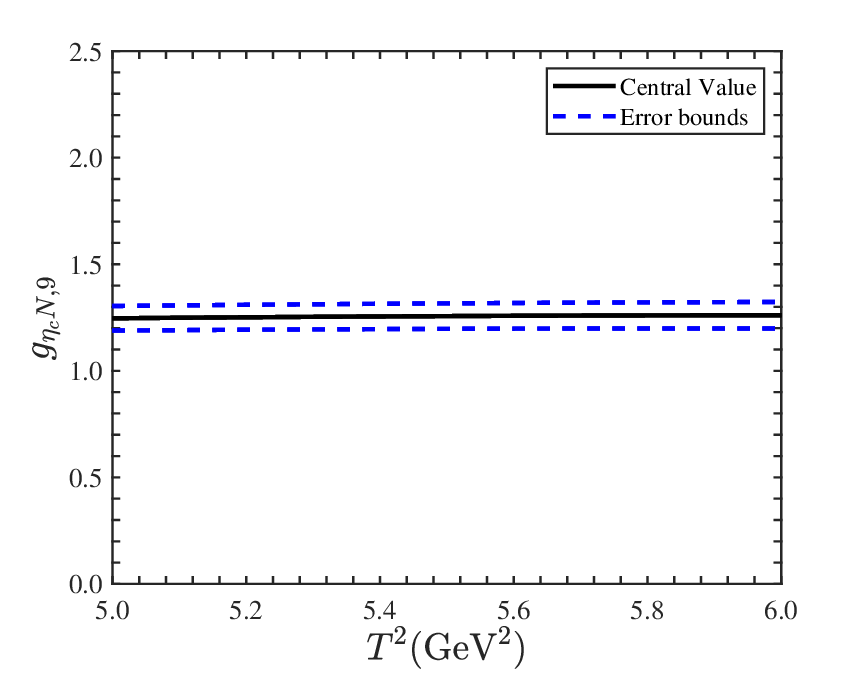}
 \includegraphics[totalheight=5cm,width=7cm]{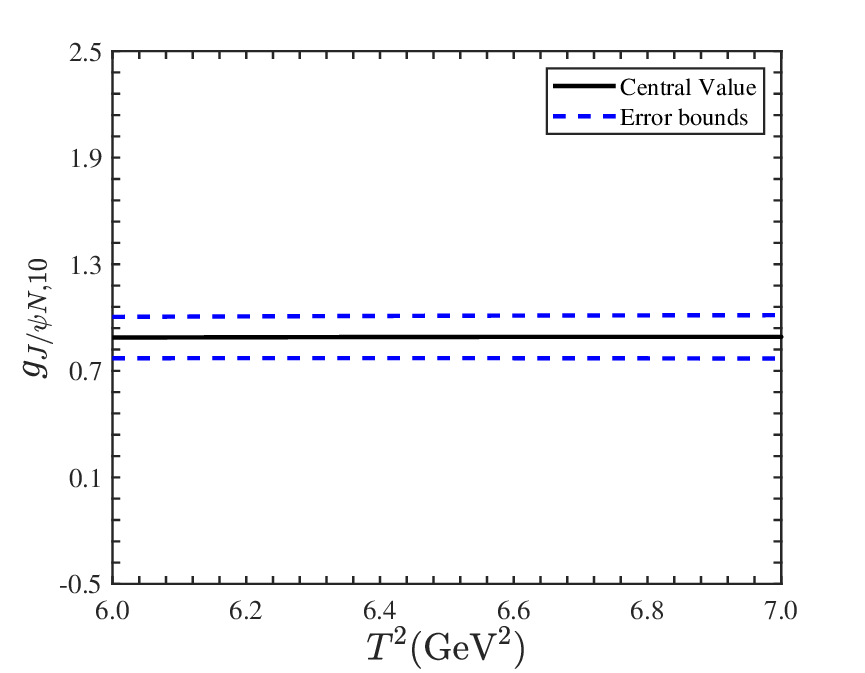}
 \caption{The hadronic coupling constants for $P_c(4457)\rightarrow \eta_cN$ (Left) and $P_c(4457)\rightarrow J/\psi N$ (Right) among the Borel windows.}\label{baryon-fig}
\end{figure}
\begin{figure}
 \centering
 \includegraphics[totalheight=5cm,width=7cm]{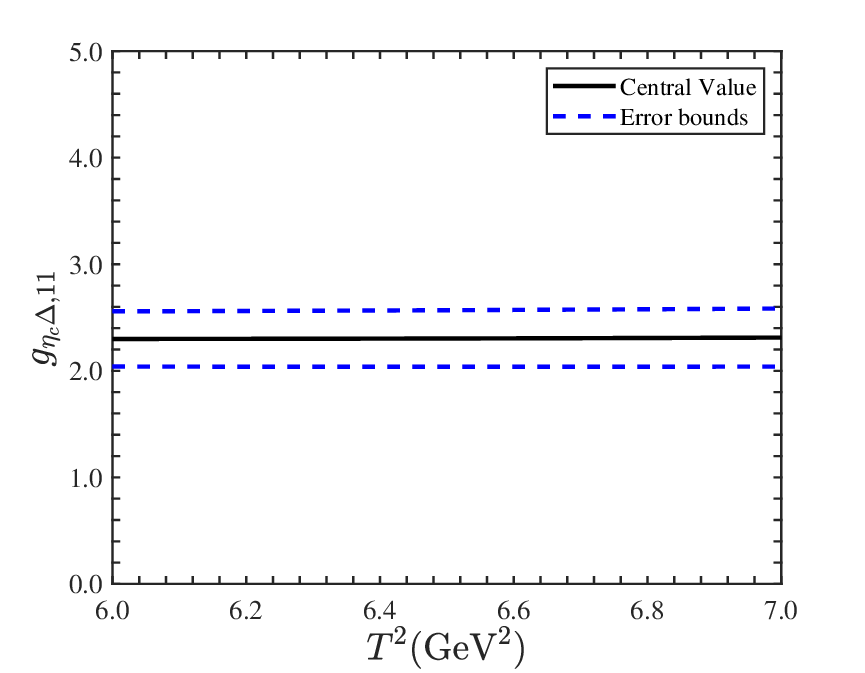}
 \includegraphics[totalheight=5cm,width=7cm]{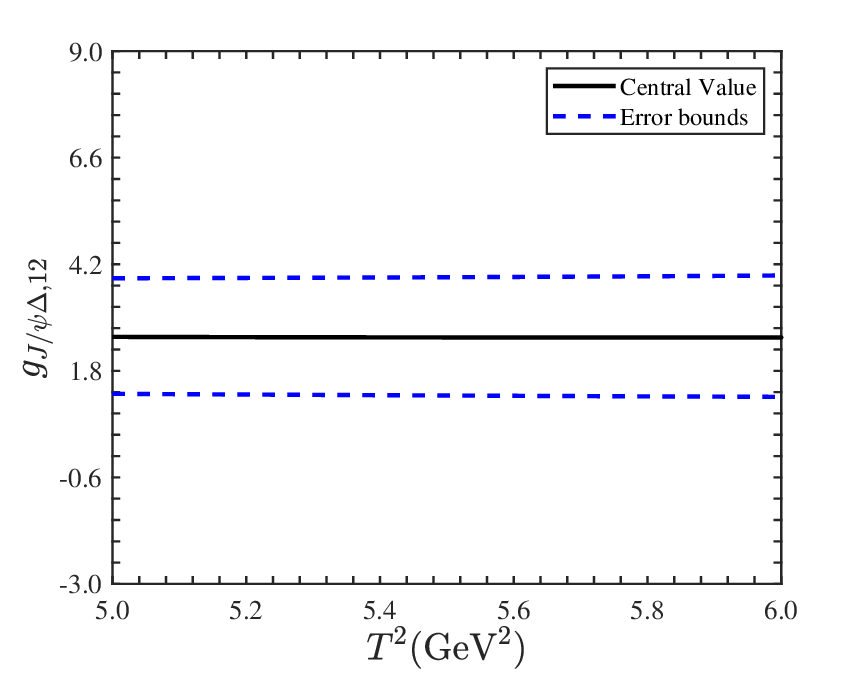}
 \caption{The hadronic coupling constants for $P_c(4620)\rightarrow \eta_c\Delta$ (Left) and $P_c(4620)\rightarrow J/\psi\Delta$ (Right) among the Borel windows.}\label{baryon-fig}
\end{figure}

\begin{figure}
 \centering
 \includegraphics[totalheight=5cm,width=7cm]{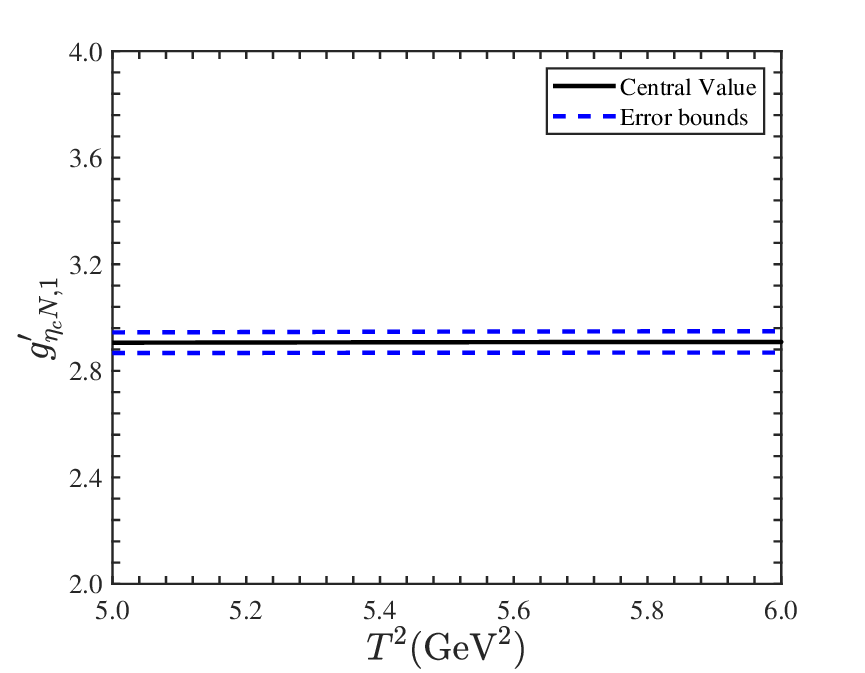}
 \caption{The hadronic coupling constant $g^{\prime}_{\eta_cN,1}$ of the decay channel $P_c(4380)\rightarrow \eta_cN$ among the Borel window for the parameter $\xi=1$.}\label{baryon-fig}
\end{figure}

\begin{figure}
 \centering
 \includegraphics[totalheight=5cm,width=7cm]{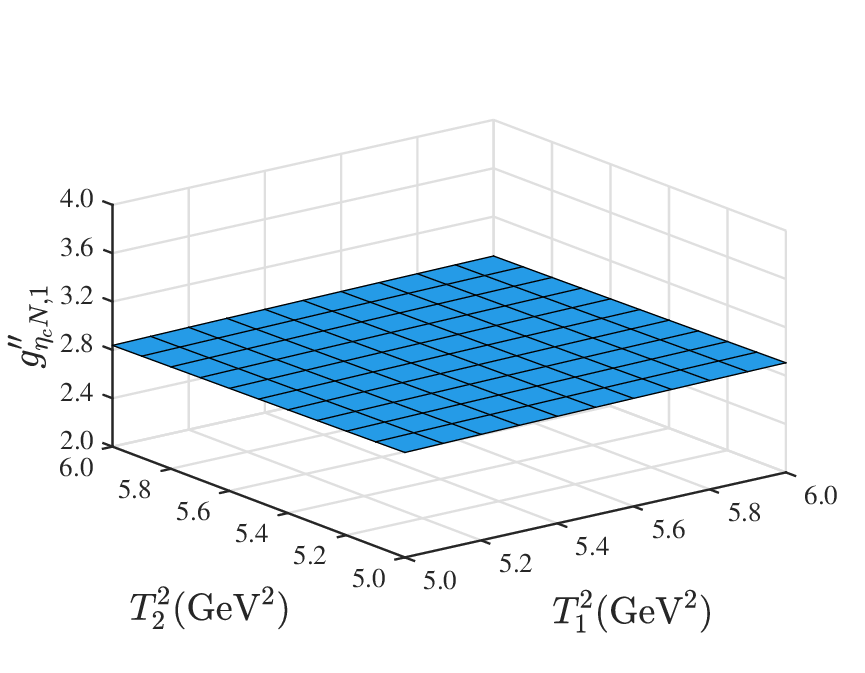}
 \caption{The 3-D graph of the hadronic coupling constant $g^{\prime\prime}_{\eta_cN,1}$ of the decay channel $P_c(4380)\rightarrow \eta_cN$ among the Borel window.}\label{baryon-fig}
\end{figure}

\begin{figure}
 \centering
 \includegraphics[totalheight=5cm,width=7cm]{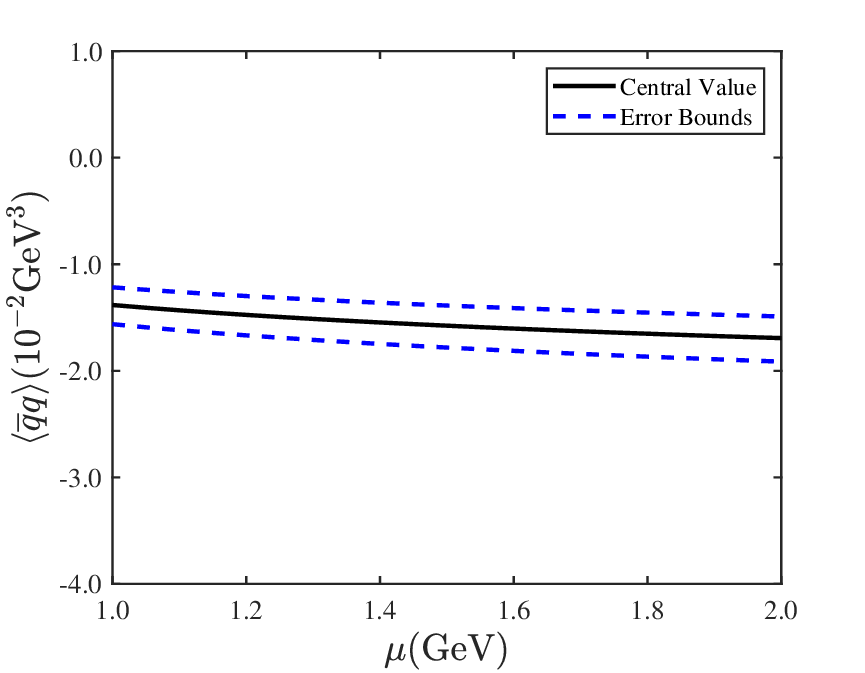}
 \includegraphics[totalheight=5cm,width=7cm]{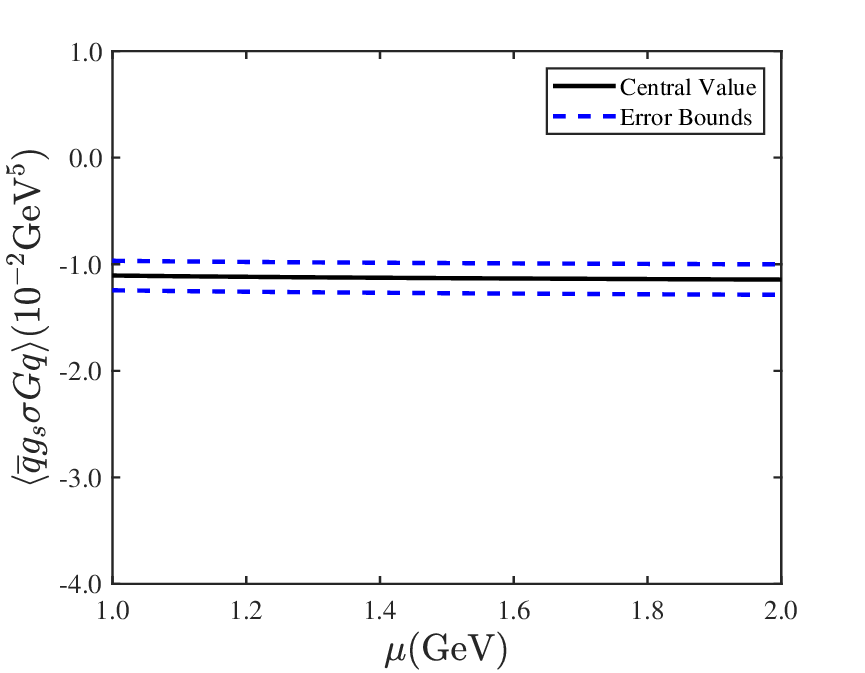}
 \caption{The energy scale dependence of the vacuum condensates $\langle\overline{q}q\rangle$ (Left) and $\langle\overline{q}g_s\sigma Gq\rangle$  (Right).}\label{energyscale-fig}
\end{figure}


\begin{thebibliography}{99}

\bibitem{RAaij1} R. Aaij et al, Phys. Rev. Lett. {\bf 115} (2015) 072001.

\bibitem{RAaij2} R. Aaij et al, Phys. Rev. Lett. {\bf 122} (2019) 222001.

\bibitem{RAaij3} R. Aaij et al, Phys. Rev. Lett. {\bf 128} (2022) 062001.

\bibitem{Myself-DcPc4312} X. W. Wang and Z. G. Wang, Chin. Phys. {\bf C48} (2024) 053102.

\bibitem{Mass-mole-JunHe} J. He, Eur. Phys. J. {\bf C79} (2019) 393.

\bibitem{Mass-mole-HXChen-2} H. X. Chen, W. Chen, X. Liu and S. L. Zhu, Phys. Rev. Lett. {\bf 115} (2015) 172001.

\bibitem{Mass-mole-HXChen} H. X. Chen, W. Chen and S. L. Zhu, Phys. Rev. {\bf D100} (2019)  051501.

\bibitem{mass-mole-WZG} Z. G. Wang,  Int. J. Mod. Phys. {\bf A34} (2019) 1950097.

\bibitem{mass-mole-MengLin} M. L. Du, V. Baru, F. K. Guo, C. Hanhart, U. G. Meissner, J.  A. Oller and Q. Wang, Phys. Rev. Lett. {\bf 124} (2020) 072001.

\bibitem{Mass-mole-LMeng} L. Meng, B. Wang, G. J. Wang and S. L. Zhu, Phys. Rev. {\bf D100} (2019)  014031.

\bibitem{Mass-mole-MingZhu} M. Z. Liu, T. W. Wu, M. S. Sanchez, M. P. Valderrama, L. S. Geng and J. J. Xie, Phys. Rev. {\bf D103} (2021) 054004.

\bibitem{Mass-JRZhang-EPJC2019} J. R. Zhang, Eur. Phys. J. {\bf C79} (2019)  1001.

\bibitem{mass-mole-Azizi} K. Azizi, Y. Sarac and H. Sundu, Chin. Phys. {\bf C45} (2021)  053103.

\bibitem{CWXiao-Pc-mole1} C. W. Xiao, J. Nieves and E. Oset, Phys. Rev. {\bf D88} (2013) 056012.

\bibitem{HXChen-Pc-mole2} H. X. Chen, E. L. Cui, W. Chen, X. Liu, T. G. Steele and S. L. Zhu, Eur. Phys. J. {\bf C76} (2016) 572.

\bibitem{MZLiu-Pc-mole3} M. Z. Liu, F. Z. Peng, M. S\'{a}nchez S\'{a}nchez and M. P. Valderrama, Phys. Rev. {\bf D98} (2018) 114030.

\bibitem{MZLiu-Pc-mole4} M. Z. Liu, Y. W. Pan, F. Z. Peng, M. S\'{a}nchez S\'{a}nchez, L. S. Geng, A. Hosaka and
M. Pavon Valderrama, Phys. Rev. Lett. {\bf 122} (2019) 242001.

\bibitem{CWXiao-Pc-mole5} C. W. Xiao, J. Nieves and E. Oset, Phys. Rev. {\bf D100} (2019) 014021.

\bibitem{Decay-mole-Sakai} S. Sakai, H. J. Jing and F. K. Guo, Phys. Rev. {\bf D100} (2019) 074007.

\bibitem{mass-penta-WZG-EPJC-16} Z. G. Wang, Eur. Phys. J. {\bf C76} (2016)  70.

\bibitem{mass-penta-WZG-IJMPA} Z. G. Wang, Int. J. Mod. Phys. {\bf A35} (2020)  2050003.

\bibitem{WZG-Pc-Penta2} Z. G. Wang, Eur. Phys. J. {\bf C76} (2016) 142.

\bibitem{WZG-Pc-Penta4} Z. G. Wang, Int. J. Mod. Phys. {\bf A36} (2021) 2150071.

\bibitem{AAli-Compact5-Pc1} A. Ali and A. Y. Parkhomenko, Phys. Lett. {\bf B793} (2019) 365.

\bibitem{RZhu-Compact5-Pc2} R. Zhu, X. Liu, H. Huang and C. F. Qiao, Phys. Lett. {\bf B797} (2019) 134869.

\bibitem{JFGiron-Compact5-Pc4} J. F. Giron, R. F. Lebed and C. T. Peterson, JHEP {\bf 05} (2019) 061.

\bibitem{FStancu-Compact5-Pc5} F. Stancu, Eur. Phys. J. {\bf C79} (2019) 957.

\bibitem{JFGiron-Compact5-Pc6} J. F. Giron and R. F. Lebed, Phys. Rev. {\bf D104} (2021) 114028.

\bibitem{Decay-mole-Zou} Y. H. Lin and B. S. Zou, Phys. Rev. {\bf D100} (2019) 056005.

\bibitem{Decay-mole-HeChen} J. He and D. Y. Chen, Eur. Phys. J. {\bf C79} (2019) 887.

\bibitem{NYalikun-Pc4} N. Yalikun, Y. H. Lin, F. K. Guo, Y. Kamiya and B. S. Zou, Phys. Rev. {\bf D104} (2021) 094039.

\bibitem{MPavon-Pc5} M. Pavon Valderrama, Phys. Rev. {\bf D100} (2019) 094028.

\bibitem{MLDu-Pc6} M. L. Du, V. Baru, F. K. Guo, C. Hanhart, U. G. Mei{\ss}ner, J. A. Oller and Q. Wang, JHEP {\bf 08} (2021) 157.

\bibitem{FZPeng-Pc7} F. Z. Peng, L. S. Geng and J. J. Xie, Phys. Rev. {\bf D111} (2025) 054029.

\bibitem{FKGuo-Review1} F. K. Guo, X. H. Liu and S. Sakai, Prog. Part. Nucl. Phys. {\bf 112} (2020) 103757 .

\bibitem{HXChen-Review2} H. X. Chen, W. Chen, X. Liu, Y. R. Liu and S. L. Zhu, Rept. Prog. Phys. {\bf 86} (2022) 026201.

\bibitem{LMeng-Review3} L. Meng, B. Wang, G. J. Wang  and S. L. Zhu, Phys. Rept. {\bf 1019} (2023) 2266.

\bibitem{ZGWang-Review4} Z. G. Wang, Front. Phys. (Beijing) {\bf 21} (2026) 016300.

\bibitem{mass-mole-WXW-SCPMA} X. W. Wang, Z. G. Wang, G. L. Yu and Q. Xin, Sci. China Phys. Mech. Astron. {\bf 65} (2022)  291011.

\bibitem{mass-mole-WXW-CPC}  X. W. Wang and Z. G. Wang, Chin. Phys. {\bf C47} (2023)  013109.

\bibitem{mass-penta-WZG-CPC} Z. G. Wang, Chin. Phys. {\bf C45} (2021)  073107.

\bibitem{Decay-mole-GJWang} G. J. Wang, L. Y. Xiao, R. Chen, X. H. Liu, X. Liu and S. L. Zhu,
Phys. Rev. {\bf D102} (2020)  036012.

\bibitem{Decay-mole-Gutsche} T. Gutsche and V. E. Lyubovitskij, Phys. Rev. {\bf D100} (2019) 094031.

\bibitem{Decay-mole-WZG-WX} Z. G. Wang and X. Wang, Chin. Phys. {\bf C44} (2020) 103102.

\bibitem{Decay-mole-HuangMQ} Y. J. Xu, C. Y. Cui, Y. L. Liu and M. Q. Huang, Phys. Rev. {\bf D102}  (2020) 034028.

\bibitem{mass-mole-WXW-IJMPA} X. W. Wang and Z. G. Wang, Int. J. Mod. Phys. {\bf A37} (2022) 2250189.

\bibitem{WZG-ZJX-Zc-Decay}   Z. G. Wang and J. X. Zhang,  Eur. Phys. J. {\bf C78} (2018)  14.

\bibitem{WZG-Y4660-Decay}    Z. G. Wang, Eur. Phys. J. {\bf C79} (2019)  184.

\bibitem{WZG-X4140-decay}     Z. G. Wang and  Z. Y. Di, Eur. Phys. J. {\bf C79} (2019)  72.

\bibitem{WZG-X4274-decay}      Z. G. Wang,    Acta Phys. Polon. {\bf B51} (2020) 435.

\bibitem{WZG-Z4600-decay}     Z. G. Wang,  Int. J. Mod. Phys. {\bf A34} (2019)  1950110.

\bibitem{WZG-Pc4312-decay-tetra}     Z. G. Wang,  H. J. Wang and  Q. Xin, Chin. Phys. {\bf C45} (2021) 063104.

\bibitem{SVZ1} M. A. Shifman, A. I. Vainshtein and V. I. Zakharov, Nucl. Phys. {\bf B147} (1979) 385.

\bibitem{SVZ2} M. A. Shifman, A. I. Vainshtein and V. I. Zakharov, Nucl. Phys. {\bf B147} (1979) 448.

\bibitem{Reinders} L. J. Reinders, H. Rubinstein and S. Yazaki, Phys. Rept. {\bf  127} (1985) 1.

\bibitem{ColangeloReview} P. Colangelo and A. Khodjamirian, hep-ph/0010175.

\bibitem{PDG}  P. A. Zyla et al, Prog. Theor. Exp. Phys. {\bf 2020}  (2020) 083C01.

\bibitem{Narison} S. Narison and R. Tarrach, Phys. Lett. {\bf  B125} (1983) 217.

\bibitem{EWZGHuang} Z. G. Wang and T. Huang, Phys. Rev. {\bf D89} (2014) 054019.

\bibitem{Becirevic} D. Becirevic, G. Duplancic, B. Klajn, B. Melic and F. Sanfilippo, Nucl. Phys. {\bf B883} (2014)
306.

\bibitem{EIoffe} B. L. Ioffe, Prog. Part. Nucl. Phys. {\bf56} (2006) 232.

\bibitem{DZGWang} Z. G. Wang, Chin. Phys. {\bf C46} (2022) 123106.

\bibitem{Myself-DcPcs4338} X. W. Wang and Z. G. Wang, Phys. Rev. {\bf D110} (2024) 014008.

\bibitem{WZG-X3872} Z. G. Wang, Phys. Rev. {\bf D109} (2024) 014017.

\bibitem{WZG-Y4500} Z. G. Wang, Nucl. Phys. {\bf B993} (2023) 116265.

\bibitem{Pcdecay0} T. J. Burns and E. S. Swanson, Eur. Phys. J. {\bf A58} (2022) 68.

\bibitem{Pcdecay01} M. Z. Liu, Y. W. Pan and L. S. Geng, Phys. Rev. {\bf D110} (2024) 114022.

\bibitem{Pcdecay02}  J. M. Xie, X. Z. Ling, M. Z. Liu and L. S. Geng, Eur. Phys. J. {\bf C82} (2022) 1061.

\bibitem{Pcdecay1} M. B. Voloshin, Phys. Rev. {\bf D100} (2019) 034020.

\bibitem{Pcdecay2} Q. Li, C. H. Chang, X. Tong, X. Z. Tan, T. H. Wang and G. L. Wang, JHEP 10 (2025) 098.

\bibitem{Pcdecay3} Fl. Stancu, Phys. Rev. {\bf D104} (2021) 054050.

\bibitem{Pcdecay4} R. Aaij et al. (LHCb), Phys. Rev. {\bf D 110} (2024) 032001.

\end{thebibliography}
\end{document}